# Penetration of Arbitrary Double Potential Barriers with Probability Unity: Implications for Testing the Existence of a Minimum Length

Yong Yang[1,2*]

1. *Key Lab of Photovoltaic and Energy Conservation Materials, Institute of Solid State Physics, HFIPS, Chinese Academy of Sciences, Hefei 230031, China.*
2. *Science Island Branch of Graduate School, University of Science and Technology of China, Hefei 230026, China.*

Quantum tunneling across double potential barriers is studied. With the assumption that the real space is a continuum, it is rigorously proved that large barriers of arbitrary shapes can be penetrated by low-energy particles with a probability of unity, i.e., realization of resonant tunneling (RT), by simply tuning the inter-barrier spacing. The results are demonstrated by tunneling of electrons and protons, in which resonant and sequential tunneling are distinguished. The critical dependence of tunneling probabilities on the barrier positions not only demonstrates the crucial role of phase factors, but also points to the possibility of ultrahigh accuracy measurements near resonance. By contrast, the existence of a nonzero minimum length puts upper bounds on the barrier size and particle mass, beyond which effective RT ceases. A scheme is suggested for dealing with the practical difficulties arising from the delocalization of particle position due to the uncertainty principle. This work opens a possible avenue for experimental tests of the existence of a minimum length based on atomic systems.

*E-mail: yyanglab@issp.ac.cn

# I. INTRODUCTION

The scenario of a minimum length ($L_{min}$) plays an essential role in the quantum theory of gravity [1-11]. Breakdown of the Lorentz invariance may happen when the real space approaches such a minimum length scale [12-20], which is generally taken to be the Planck length ($l_P = \sqrt{\frac{\hbar G}{c^3}} \sim 1.6 \times 10^{-35}\ m$, $\hbar$ is the reduced Planck's constant, $G$ is the gravitational constant and $c$ is the speed of light) [7]. Despite numerous efforts [1-20], the question remains open regarding the existence of such a minimum length [7, 21-23]. Since such a length scale is well below the lower bound of spatial resolution achieved by state-of-the-art instruments such as LIGO ($\sim 10^{-19}\ m$) [24, 25], it is a great challenge for experimental verification. Here, we show the possibility of tackling this problem by investigations of quantum tunneling across double potential barriers.

Quantum tunneling [26] is a classically forbidden phenomenon in which a particle passes through a potential barrier higher than the energy it possesses. In the early years of quantum mechanics, the theories based on quantum tunneling explain some puzzles of experimental observations like the thermionic and field-induced emission of electrons from metal surfaces [26], and the alpha decay of heavy nuclei [26]. In the ensuing decades, researches on the quantum tunneling of electrons in condensed matter have led to fruitful discoveries [27-33], and enabled important inventions such as the scanning tunneling microscope (STM) [34] and tunneling diodes [35-37]. Since the pioneering works by Tsu, Esaki and Chang [31-33], double barriers have received a lot of attention while studying electron transport in heterostructures [37, 38]. Resonant tunneling (RT) typically takes place in double-barrier systems, in which the incident electrons may pass through the barriers without being reflected, i.e., with a transmission probability of 100%. Such a behavior is due to the coherent interference of electron waves which cancel the reflected waves and enhance the transmitted ones, analogous to the resonant transmission through a Fabry-Perot etalon in optics. Typical inter-barrier spacing of the devices based on RT is several tens of angstroms (Å), matching the de Broglie wavelengths of electrons. In recent decades, the phenomena of RT in mesoscopic and nanoscale structures continue to attract interests of research [39-43].

Historically, RT of electrons was considered to gain experimental evidence from the negative differential resistance (NDR) found in the current-voltage (*I-V*) curves [32, 35, 37]. Later, alternative mechanism was suggested for NDR, namely, sequential tunneling in which the phase memory of electron wave functions is lost due to inelastic scattering [37, 44-51]. It was argued that resonant (coherent) tunneling is a prerequisite for sequential tunneling [51]. The effects of external electric field, inelastic scattering, and the repulsive interactions between electrons on RT were also studied [38, 52, 53]. In spite of these efforts, consensus on the underlying physics is yet to be reached. There are still large discrepancies between theory and the measured *I-V* curves (e.g., peak-to-valley ratio). The gap originates partly from the fact that, in calculations related to experiments the simplest rectangular barriers (or their variants) are adopted, which usually differ significantly from the true barriers felt by electrons.

To resolve the puzzles, *exact theoretical description of the conditions for RT* across double barriers is highly desired. For the simplest rectangular double barriers, exact mathematical relation of energy and geometric conditions has been established [38, 54, 55]. For the more general and realistic situation where double barriers are of arbitrary shapes, aside from the semi-classical approach [38], full quantum level description of the RT conditions is still lacked. It is generally accepted that RT takes place when the energy of incident particle matches the energy levels of the quasi-bound states within the potential well in-between the two barriers [37, 38, 44, 49, 51, 52]. In principle, this applies to electrons as well as the massive particles like protons, atoms and molecules. Recent simulations have shown the RT of H and He atoms across small double barriers, with the barrier height $E_b \sim 0.2$ eV [56, 57] and $\sim 0.02$ eV [58], respectively. However, when a particle tunnels across arbitrarily-shaped double-barriers, it is unclear how the level-match condition can be reached, and rigorous theoretical descriptions remain elusive.

In this paper, we revisit this topic in double-barrier systems consisting of equal barriers of arbitrary geometries. With the assumption of a continuously varied inter-barrier spacing (equivalently, $L_{min} = 0$), it is rigorously proved that quantum tunneling through the double-barrier system with a probability of unity can always happen (i.e., RT) when the inter-barrier spacing is appropriately chosen. Exact mathematical relation for RT is established. At the presence of a nonzero $L_{min}$ ($L_{min} > 0$), the inter-barrier spacing varies discontinuously, which sets upper bounds for the barrier heights and particle mass, above which no RT may happen. The results are demonstrated by the tunneling of electrons, protons, and some typical bosons. Practically possible scheme is therefore provided for experimental tests of the existence of a nonzero minimum length.

The rest of this paper is organized as follows. Section II presents the analytic and numerical results on RT, with examples of typical particles like electrons, protons and some bosons. The connection between RT and the continuity of real space is revealed. The constraints set by the existence of a minimum length, the practical obstacles due to the uncertainty principle and plausible solutions are presented. We conclude in Sec. III with discussions on the impacts and future opportunities inspired by this work.

## II. RESULTS AND DISCUSSIONS

We begin in Part A of this section by performing general analysis on the transmission properties of quantum particles across double barriers of arbitrary shapes, and prove a theorem which establishes the mathematical condition for resonant tunneling (RT). Part B provides analyses on two typical models — rectangular and parabolic double barriers. The quantum tunneling of electrons and protons are studied and compared, with emphasis on the differences between resonant and sequential tunneling. Based on the results of Parts A and B, we show in Part C the upper bounds of barrier heights for RT set by the Planck length. In Part D, the fundamental limits put by the uncertainty principle are studied and possible solution to position delocalization of the incident particles is suggested.

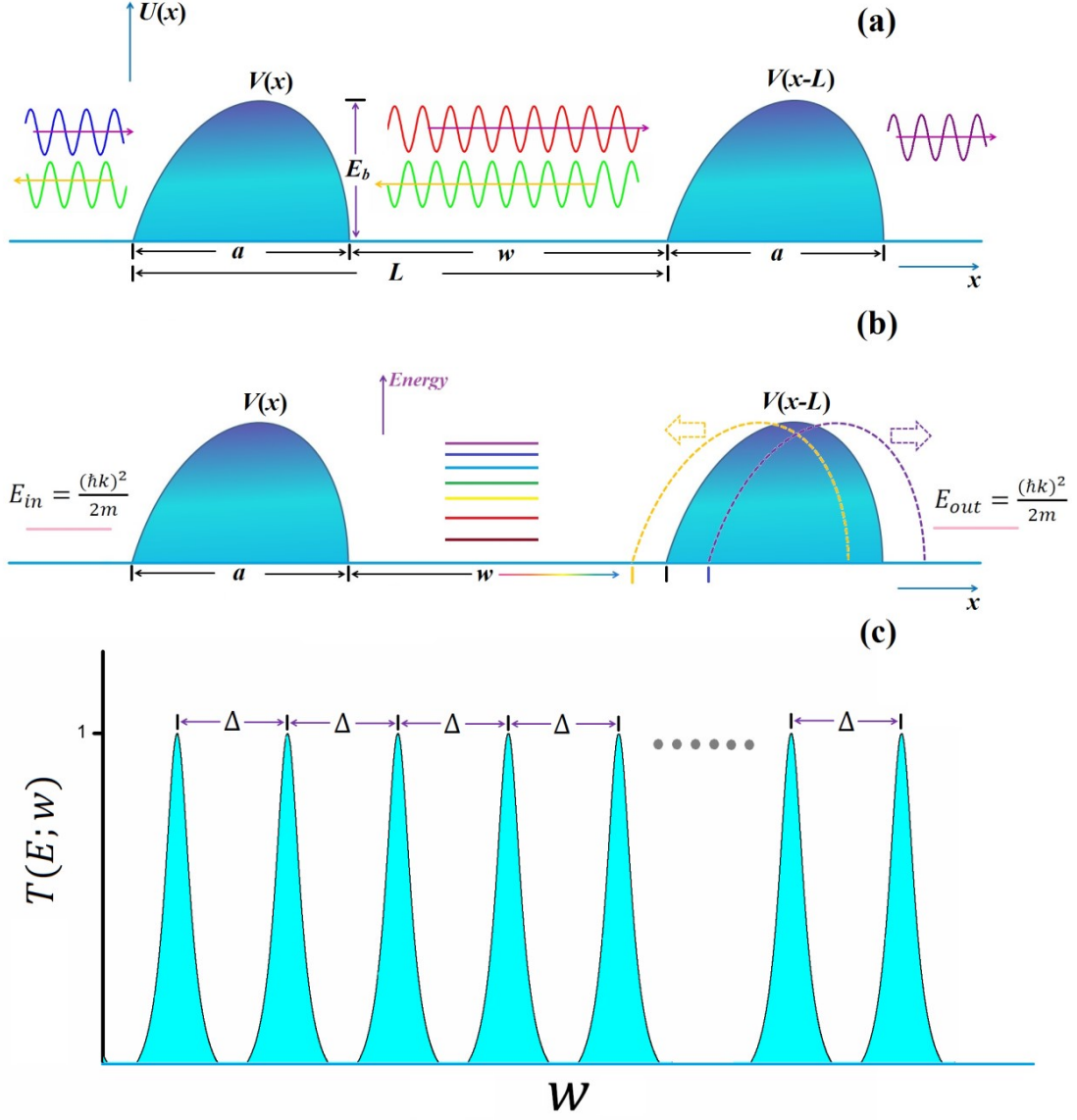

**FIG. 1**. Schematics of resonant tunneling (RT) across double barriers. **(a)** Quantum interference of the incident and reflected matter waves; **(b)** Modulation of the energy levels of the quasi-bound states in-between the two barriers, by varying the inter-barrier spacing $w$; **(c)** RT spectrum as a function of $w$ with a period of $\Delta$ $(=\frac{\pi}{k})$.

## A. GENERAL ANALYSES ON ARBITARY DOUBLE BARRIERS

Generally, double-barriers consist of two identical or different single barriers, which are respectively referred to as homo-structured and hetero-structured hereafter. Unless otherwise stated, the double-barrier considered here is homo-structured in one-dimensional space, as shown in Fig. 1(a), with a barrier height $E_b$ and barrier width $a$ for each. Our analyses are based on the transfer matrix method, a powerful technique for studying the transmission properties in finite systems [31, 59-62]. For the propagation of a quantum particle across a single barrier $V(x)$ with compact support (the intrinsic property of physical interactions), the transmitted and reflected amplitudes ($A_L$, $B_L$; $A_R$, $B_R$) of the wave functions ($\psi_L$, $\psi_R$) may be related by a transfer matrix (denoted by $M$) as follows [38, 56, 59-62]:

$$\begin{pmatrix} A_R \\ B_R \end{pmatrix} = M \begin{pmatrix} A_L \\ B_L \end{pmatrix} \equiv \begin{pmatrix} m_{11} & m_{12} \\ m_{21} & m_{22} \end{pmatrix} \begin{pmatrix} A_L \\ B_L \end{pmatrix}. \qquad (1)$$

The incoming wave function (with incident energy $E$) is $\psi_L = A_L e^{ikx} + B_L e^{-ikx}$, and the outgoing wave function is $\psi_R = A_R e^{ikx} + B_R e^{-ikx}$, where $k = \sqrt{2mE/\hbar^2}$, $m$ is the particle mass. The determinant $|M| = 1$, for systems where time-reversal symmetry preserves, and the transmission coefficient is given by [56] $T = \frac{1}{|m_{11}|^2} = \frac{1}{|m_{22}|^2}$. In general, the matrix elements $m_{ij}$ ($i, j$ = 1, 2) are complex numbers and obey the conjugate relations [56, 59-62] of $m_{11}^* = m_{22}$, and $m_{12}^* = m_{21}$. For a homo-structured double-barrier with an inter-barrier spacing $w$, the following theorem holds:

***Theorem.*** – For any $E < E_b$, the transmission coefficient (tunneling probability) across a homo-structured double-barrier $T_{DB}(E; w) = 1$ at $w = w_n = \frac{n\pi}{k} - \frac{\pi + \theta + 2ka}{2k}$, where $\theta = \arg(m_{11}{}^2)$, $n$ (referred to as resonance number) belongs to integers.

***Proof.*** – The updated transfer matrix for a single barrier $V(x)$ translated by a distance $L = a + w$, $V(x-L)$, is given by [57, 63]

$$M(L) = \begin{pmatrix} m_{11} & m_{12}e^{-i2kL} \\ m_{21}e^{i2kL} & m_{22} \end{pmatrix} = \begin{pmatrix} m_{11} & m_{12}e^{-i2k(a+w)} \\ m_{21}e^{i2k(a+w)} & m_{22} \end{pmatrix}. \qquad (2)$$

The transfer matrix for the double-barrier ($U(x) = V(x) + V(x-L)$) is therefore [56]

$$M_{DB} = M(L) * M = \begin{pmatrix} m_{11} & m_{12}e^{-i2k(a+w)} \\ m_{21}e^{i2k(a+w)} & m_{22} \end{pmatrix} \begin{pmatrix} m_{11} & m_{12} \\ m_{21} & m_{22} \end{pmatrix}. \qquad (3)$$

The diagonal matrix element describing the transmission properties, $(M_{DB})_{11}$, is explicitly calculated to be $(M_{DB})_{11} = m_{11}{}^2 + m_{12}m_{21}e^{-i2k(a+w)}$. Let $Z = m_{11}{}^2 \equiv |Z|e^{i\theta}$, $\phi = 2k(a+w)$, $m_{12}m_{21} = |m_{12}|^2 = R$, with $i$ the imaginary unit, and the angle $\theta = \arg(Z)$; the determinant, $|M| = 1 = |m_{11}|^2 - |m_{12}|^2 = |Z| - R$ gives that $|Z| = 1 + R$, then $(M_{DB})_{11} = (1 + R)e^{i\theta} + Re^{-i\phi} = e^{i\theta} + Re^{i\theta}(e^{-i(\phi+\theta)} + 1)$. When $e^{-i(\phi+\theta)} = -1$, i.e., $\phi + \theta = (2n - 1)\pi$, with $n$ being integers, one has $(M_{DB})_{11} = e^{i\theta}$. It follows that the transmission coefficient $T_{DB}(E; w) = \frac{1}{|(M_{DB})_{11}|^2} = 1$, which corresponds to RT. Using the condition $\phi + \theta = (2n - 1)\pi$, one has $2k(a + w) + \theta = (2n - 1)\pi$, and consequently, $w = \frac{n\pi}{k} - \frac{\pi + \theta + 2ka}{2k} \equiv w_n$. This completes the proof of the theorem.

It should be stressed here that the proof inherently includes the precondition that the inter-barrier spacing $w$ varies continuously ($L_{min} = 0$) such that the angle $\phi$ can have any desired values to satisfy the RT condition. The theorem points to the possibility of *penetration of arbitrarily large (but finite) potential barriers by low-energy particles with a probability of unity*. For a quantum particle with incident energy $E$, it can completely tunnel across a homo-structured double barrier of height $E_b$ when the inter-barrier spacing equals $w_n$ described above, even in the case $E << E_b$. In addition, one sees that the barrier-barrier separations ($w_n$) for RT are solely determined by the parameters ($\theta$, $a$) describing the transmission of single barriers. Physically, the onset of RT is due to the presence of quasi-bound states in-between the two barriers whose energy levels match that of the incident particles [37, 38, 44, 49, 51, 52]. A direct consequence is that, any quasi-bound energy levels ($E \leq E_b$) can be realized within the potential well set by the two barriers via simply tuning the inter-barrier spacing, as illustrated in Fig. 1(b). Moreover, from its mathematical expression, one sees that $(M_{DB})_{11}$ is the periodic function of $w$, with a period of $\tau = \frac{\pi}{k}$. For a fixed $E$, the tunneling probability $T[E; w]$ displays periodic variations with $w$, showing comb-like structures with the resonance peaks positioned at $L_n = a + w_n$, and the distance between any two neighboring peaks is $\Delta = w_n - w_{n-1} = \frac{\pi}{k}$ (Fig. 1(c)). The value of $\Delta$ is just half the de Broglie wavelength of the incident matter wave, indicating the key role of phase factor and quantum interference. Finally, the mathematical expression of $w_n$ implies that there could be infinitely many resonance peaks in free space. The theorem applies rigorously to point-like particles such as electrons [64, 65]. We go on to show that the theorem holds valid for finite-size particles. For a particle incident along the *x* direction, the effects due to finite size are determined by the distribution ρ(*x*) of the physical quantity (e.g., charge) that the barriers (e.g., Coulomb interactions) arise from. Generally, a normalized ρ(*x*) is subjected to the constraint $\int_{x_1}^{x_2} \rho(x)dx = 1$, with $x_1$, $x_2$ being the coordinates of two edge points. By defining the weight averaged center $x_c = \int_{x_1}^{x_2} x\rho(x)dx$, one can introduce the internal coordinate τ with reference to $x_c$, which is translational invariant: τ = *x* - $x_c$, $\tau_1$ = $x_1$ - $x_c$, $\tau_2$ = $x_2$ - $x_c$, and $\int_{\tau_1}^{\tau_2} \rho(\tau)d\tau = 1$. Given that $V_0(x)$ is the potential barrier felt by a point-like particle, the true barrier felt by a finite-size incident particle is given by $V(x) = \int_{\tau_1}^{\tau_2} V_0(x)\rho(\tau)d\tau$. The double-barrier felt by the particle is therefore changed from $U_0(x)$ (= $V_0(x)$ + $V_0(x-L)$) to $U(x)$ (= $V(x)$ + $V(x-L)$), with the variation δ$U(x)$ = $U(x)$ - $U_0(x)$. As a result, the finite-size effects are self-consistently incorporated into the new double-barrier $U(x)$. It is straightforward that the theorem holds and the incident particle behaves as a point-like particle with the coordinate $x_c$ during the process of tunneling across the new double-barrier. The results may be readily extended to two- or three-dimensional systems in that the interaction potentials along the direction of propagation are equivalently described by some effective double barriers, by considering the translational invariance in the plane perpendicular to the direction of propagation.

## B. TUNNELING ACROSS TYPICAL DOUBLE BARRIERS: RESONANT TUNNELING VERSUS SEQUENTIAL TUNNELING

For homo-structured rectangular double barriers, analytic expressions of $w_n$ are available, which enable in-depth understanding of the physics of RT. The matrix element $m_{11}$ describing the transmission across single rectangular barrier (barrier height $V_0$) may be expressed as follows (Appendix A):

$$m_{11} = 2\gamma e^{-ika}[i(k^2 - \beta^2)\sinh(\beta a) + 2\beta k\cosh(\beta a)], \quad (4)$$

where $k = \sqrt{2mE/\hbar^2}$, $\beta = \sqrt{2m(V_0 - E)/\hbar^2}$, $\gamma = \frac{1}{4\beta k}$. Eq. (4) may be reduced to

$$m_{11} = 2\gamma e^{-ika} \times \sigma e^{i\alpha} = 2\gamma\sigma e^{i(\alpha-ka)}, \qquad (5)$$

where $\sigma = \sqrt{A^2 + B^2}$, $A = (k^2 - \beta^2)\sinh(\beta a)$, $B = 2\beta k \cosh(\beta a)$, and the angle $\alpha = \arctan\left(\frac{A}{B}\right)$. Therefore, $m_{11}^2 = \frac{(A^2+B^2)}{4\beta^2 k^2} e^{i2(\alpha-ka)}$. Using the theorem stated above, the angle $\theta = 2(\alpha - ka)$, and then $\theta + 2ka = 2\alpha$. The inter-barrier spacing is given by $w_n = \frac{n\pi}{k} - \frac{\pi + 2\alpha}{2k}$. It follows that $2kw_n = (2n - 1)\pi - 2\alpha$, and one arrives at the equality: $\tan(2kw_n) = \frac{\delta \tanh(\beta a)}{1 - \frac{1}{4}\delta^2 \tanh^2(\beta a)}$, where $\delta \equiv \left(\frac{\beta}{k} - \frac{k}{\beta}\right)$. Alternatively, this equality can be obtained by direct calculation of the squared norm of diagonal element $|(M_{DB})_{11}|^2$, a function of inter-barrier spacing $w$: The minimum of $|(M_{DB})_{11}|^2$ leads to RT (Appendix B). The equality for $\tan(2kw_n)$ is in line with Ref. [55], which was derived in a different way. In the special case when the incident energy is half the barrier height ($E = 0.5V_0$), $\beta = k$, the angle $\alpha = 0$, one obtains a simplified relation that $2kw_n = (2n - 1)\pi$, and $w_n = \frac{(n-1/2)\pi}{k} = \left(n - \frac{1}{2}\right)\left(\frac{\lambda_d}{2}\right)$, with $\lambda_d = \frac{2\pi}{k}$, is the de Broglie wavelength. In another special case when $k \ll \beta$ and $\beta a \gg 1$, i.e., the incident energy is far below the barrier height, one has $\alpha \cong -\frac{\pi}{2} + \frac{k}{2\beta}$ and $kw_n \cong n\pi - \frac{k}{2\beta}$, $w_n \cong \frac{n\pi}{k} - \frac{1}{2\beta}$. In both situations, the value of $w_n$ is independent of the barrier width.

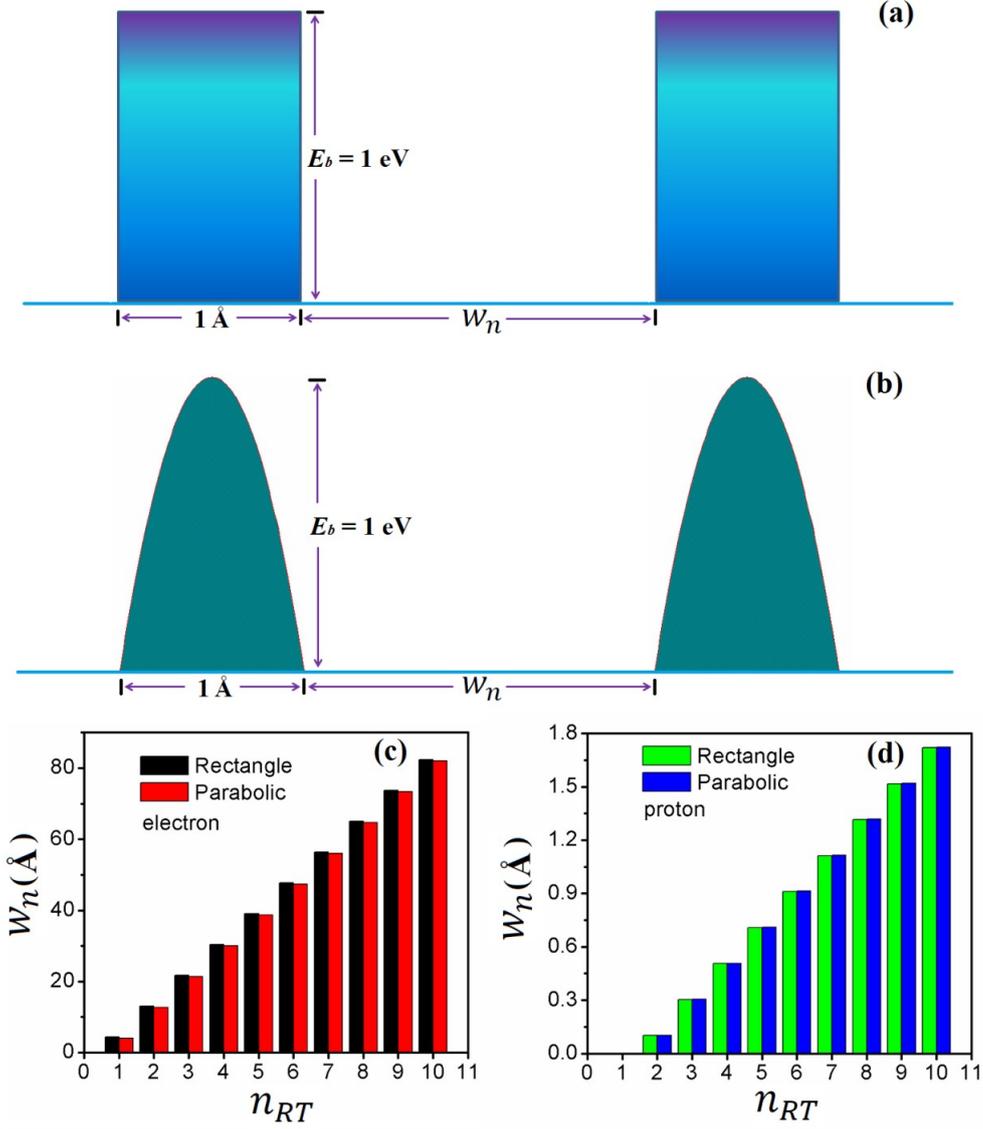

**FIG. 2.** Schematics of rectangular **(a)** and parabolic double barriers **(b)**. The inter-barrier spacing of RT ($w_n$), as a function of resonance number ($n_{RT}$), for electron **(c)** and proton **(d)** across the double barriers, at incident energy $E = 0.5$ eV.

The results are applicable to electrons and other massive quantum particles. However, demonstration of the quantum interference effects leading to RT would be much more challenging due to the large difference in particle masses and the corresponding de Broglie's wavelengths. Here, we perform systematic investigations on the RT characteristics of electrons and protons through two model systems: rectangular and parabolic double barriers (Fig. 2). Compared to the analytic expressions for rectangular barriers, the transfer matrices for parabolic barriers are evaluated numerically [56, 57]. Figures 2(c)-(d) shows the calculated $w_n$ for the RT of electrons and protons across the two types of double barriers, as a function of the resonance numbers ($n_{RT}$). For the same $n_{RT}$, the $w_n$ of electrons is much larger than that of proton owing to smaller mass. The different geometries of the potential barriers (rectangular vs parabolic) are reflected by the slight differences of $w_n$. Despite the differences, the overall comparable magnitudes of the two sets of $w_n$ indicate that rectangular double barriers may serve as approximations for qualitative description of some smoothly varying double barriers with regular geometries.

At fixed energy $E$, the tunneling probability varies periodically with inter-barrier spacing $w$. We have further studied such characteristics in case of electrons and protons tunneling through rectangular double barriers. Figures 3(a-b) show the transmission of electrons, at varying $w$ for $E$ = 0.03 eV and 0.5 eV. The effects of incident energy on the tunneling spectrum, $T(E; w)$, are clearly seen. Higher energy not only results in smaller period of oscillation ( $\tau = \frac{\pi}{k}$ ), but also smaller peak-to-valley ratio. The resonance number can extend to very large integers, as long as the perturbation from the environment is negligible and the coherence of wave functions is maintained. To show the role of coherence, we have studied the energy-dependent tunneling probability $P(E)$ of electrons at a fixed inter-barrier spacing ($w \sim 10\mu m$). For resonant (coherent) tunneling, the quantity $P(E)$ (= $T(E; w)$) drops quickly with small deviations from the resonant energy level, $E_{RT}$. For sequential tunneling, in which the phase coherence is destroyed in a two-step tunneling process, the quantity $P(E)$ is simply product of the transmission coefficient across each single barriers: $P(E) = T_1(E) \times T_2(E) = T_1^2(E)$. Around the resonant energy $E_{RT}$ (Figs. 3(c-d)), the probability of sequential tunneling ($P_{ST}$) changes smoothly with energy, and is significantly smaller than unity at low incident energies.

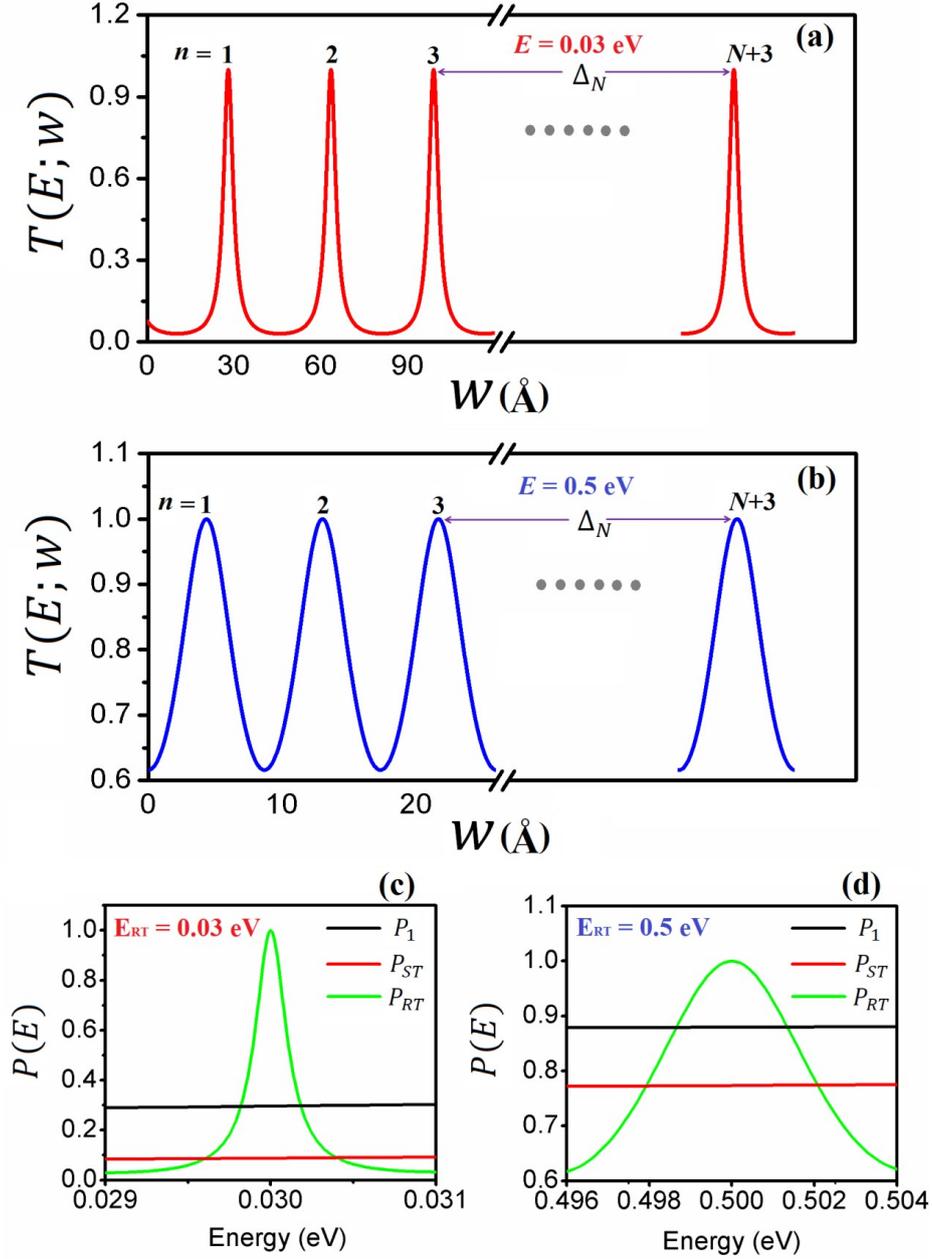

**FIG. 3.** Tunneling spectrum of electrons across the rectangular double-barrier shown in Fig. 2(a). RT at resonance level $E_{RT}$ = 0.03 eV **(a)** and 0.5 eV **(b)**, as a function of inter-barrier spacing $w$. Panels **(c-d)**: Energy dependence of tunneling probability at fixed $w$, around $E_{RT}$ = 0.03 eV (panel **c**, $w$ = 100008.49 Å) and 0.5 eV (panel **d**, $w$ = 100991.45 Å). The data lines labeled by $P_1$, $P_{ST}$ and $P_{RT}$ correspond to tunneling through single barrier, sequential and resonant tunneling through double-barrier, respectively.

For protons, more radical differences encounter. Shown in Fig. 4(a), is the tunneling spectrum of protons at $E = E_b/2 = 0.5$ eV. The periodically repeated isolated lines imply much narrower resonant peaks with comparison to electrons. The enlarged structures of one resonant peak are shown in Fig. 4(b). Around the RT peaks, the squared norm of transfer matrix element may be expressed as follows (Appendix C):

$$|(M_{DB})_{11}|^2 \cong 1 + \sinh^2(2ka) \times (k\Delta w)^2 \equiv 1 + \Delta|M_{11}|^2_{\Delta w} , \qquad (6)$$

where $\Delta w$ is the deviation from the peak position $w_n$. When $\Delta|M_{11}|^2_{\Delta w} = 1$, $T(E; w) = 0.5$, and one has

$$|\Delta w| = \frac{1}{k\sinh(2ka)} . \qquad (7)$$

It follows that the term $2|\Delta w|$ is the full width at half maximum (FWHM) of the resonant peaks. For the double-barrier (Fig. 2(a)) considered here, it turns out that $|\Delta w| \cong 4.235 \times 10^{-15}$ Å. When the deviation $\Delta w \sim 10^{-13}$ Å, the tunneling probability drops quickly to $T(E; w) \sim 10^{-3}$, in good agreement with the results presented in Fig. 4(b). In general, given that $w$ is an approximate value to $w_n$, one can determine the significant digits of $w$ by designating a deviation $\Delta w$ when $|w - w_n| \leq \Delta w$ such that $T(E; w) \geq 1 - \delta P$, where $\delta P$ ($0 < \delta P < 1$) is the tolerance of decrease in tunneling probability at which significant tunneling (effective RT, events measurable in experiment) is maintained. Furthermore, using the proof of the theorem, we find that for arbitrarily homo-structured double barriers, the deviation $\Delta w$ at the tolerance $\delta P$ may be given by (Appendix D):

$$\Delta w = \frac{1}{2k}\sqrt{\frac{1}{R(1+R)} \times \frac{\delta P}{1-\delta P}} , \qquad (8)$$

where $R = |m_{12}|^2$, $k$ and $\delta P$ are defined as above. Here we focus on the case that $\delta P = 0.5$, which yields the FWHM ($= 2\Delta w$). Indeed, the high sensitivity on barrier positions has been revealed by studies on resonant tunneling transducers [66], which are based on electronic double-barrier systems. Later, it was theoretically proposed that a tunneling electromechanical transducer may be employed to dynamically detect the Casimir forces between two conducting surfaces [67].

Such ultrahigh sensitivity on tunneling parameters is also found for the RT energies. Figure 4(c) compares the tunneling of protons across single and double barriers at a fixed inter-barrier spacing ($w \sim 20$ Å). Near resonance, $P(E)$ descends drastically from 1 to $\sim 10^{-11}$ by a tiny shift of $\varepsilon = 10^{-10}$ eV from $E_{RT}$. At the vicinity of $E_{RT}$, the dependence of $|(M_{DB})_{11}|^2$ with deviation $\Delta E$ is given by (Appendix C):

$$|(M_{DB})_{11}|^2 \cong 1 + \sinh^2(2ka) \times \left(\frac{kw}{2}\right)^2 \times \left(\frac{\Delta E}{E}\right)^2 \equiv 1 + \Delta|M_{11}|^2_{\Delta E}.$$ The FWHM at energy scale is therefore obtained when $\Delta|M_{11}|^2_{\Delta E} = 1$, and $\left|\frac{\Delta E}{E}\right| = \frac{2}{(kw) \times \sinh(2ka)}$. In our case, $\left|\frac{\Delta E}{E}\right| \approx 4.235 \times 10^{-16}$. When the energy broadening $\Delta E = \varepsilon = 10^{-10}$ eV, $\left|\frac{\Delta E}{E}\right| = 2 \times 10^{-10}$, $|(M_{DB})_{11}|^2 \cong 2.23 \times 10^{11}$, $P(E) = |(M_{DB})_{11}|^{-2} \approx 10^{-11.35}$, compares well with the numerical results. Without resonance, the probability of a two-step tunneling, i.e., sequential tunneling decreases by more than 25 orders of magnitude (Fig. 4(c)). The sharp contrast distinguishes RT from sequential tunneling. For the more generalized case, the allowed energy broadening $\Delta E$ may be calculated as follows (Appendix D):

$$\left|\frac{\Delta E}{E}\right| = \frac{1}{k(a+w)} \sqrt{\frac{1}{R(1+R)}} \times \frac{\delta P}{1-\delta P} \quad . \tag{9}$$

In the case of $R \gg 1$ (large reflection), for instance, tunneling through large barriers or tunneling by massive particles, Eq. (9) is reduced to $\left|\frac{\Delta E}{E}\right| \cong \frac{1}{k(a+w)R} \sqrt{\frac{\delta P}{1-\delta P}}$. For a single barrier $V(x)$, the reflection and tunneling probabilities (Appendix D) are related to $R$ by $|r|^2 = R|t|^2 \equiv RT_1(E)$, subjected to the condition $|r|^2 + |t|^2 = 1$. It is straightforward that $R = \frac{1}{|t|^2} - 1 \cong \frac{1}{|t|^2} = \frac{1}{T_1(E)}$ when $R \gg 1$, where $T_1(E)$ is the tunneling probability across a single barrier.

As seen from Fig. 4(c), at the absence of phase coherence, the incident protons will be nearly completely reflected by a single barrier. On the contrary, when the inter-barrier spacing equals $w_n$ and phase coherence is maintained, the protons penetrate the two barriers with a probability of unity. Such effects are schematically illustrated in Fig. 5. The key role of quantum interference is demonstrated. Experimental verification may be carried out using atomically thin membranes, which have potential applications as proton sieve filters. Generally, the variation step ($\Delta l$) of the inter-barrier spacing $w_n$ required by RT should be the order of magnitude of $\Delta w$ studied above, and no less than the minimum length (i.e., $\Delta l \sim \Delta w \geq L_{min}$) such that effective RT can be reached by tuning the inter-barrier spacing. This is the topic of next subsection.

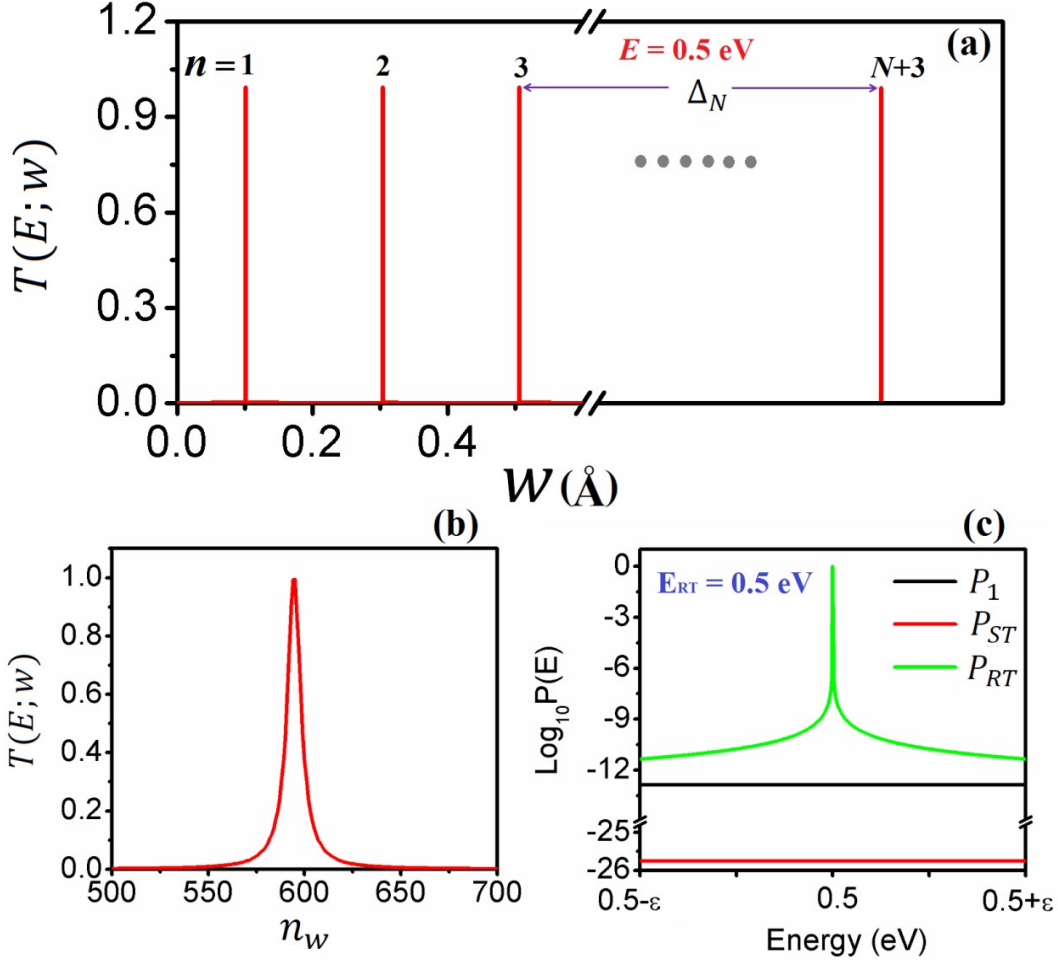

**FIG. 4.** Tunneling spectrum of protons across the rectangular double-barrier shown in Fig. 2(a). Resonance at $E_{RT} = 0.5$ eV **(a)**. Panels **(b-c)**: Variations of tunneling probability with respect to small deviations from the RT parameters: **(b)** Inter-barrier spacing $w = (n_w - n_p) \times \Delta l + w_n$, $n_p = 596$ and $\Delta l = 10^{-15}$ Å; **(c)** Incident energy at the vicinity of $E_{RT}$, for $w = 20.137016632763302$ Å, and the energy deviation $\varepsilon = 10^{-10}$ eV. All digits of $w$ are meaningful.

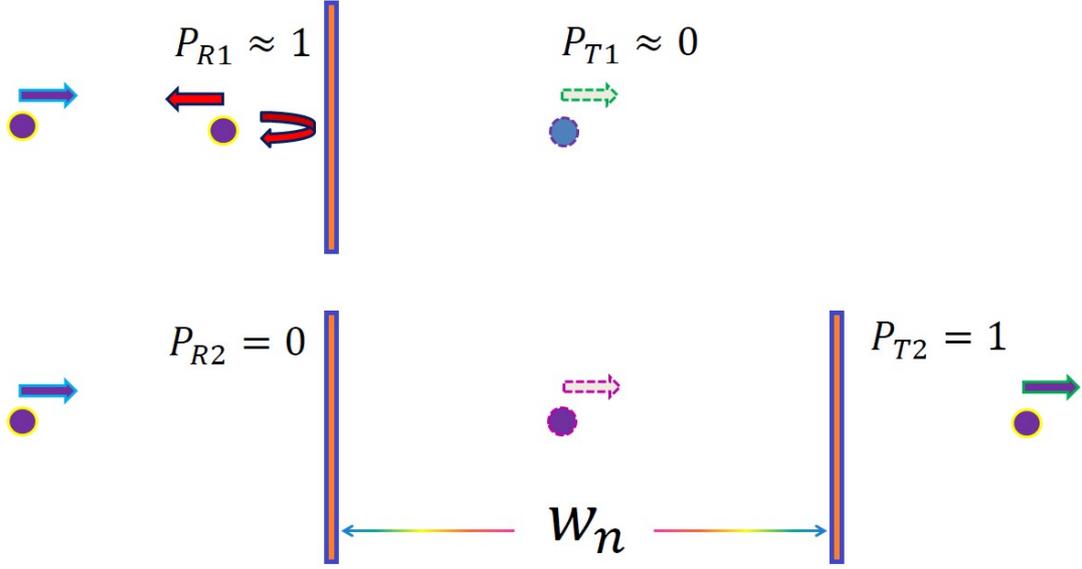

**FIG. 5.** Schematics of quantum tunneling of protons across single barrier (upper panel) and double barriers (lower panel) at the presence of RT. The probability of reflection is denoted by $P_{Ri}$ and tunneling by $P_{Ti}$, $i = 1, 2$.

To reveal the effects of asymmetry, we have studied the transmission properties of hetero-structured rectangular double-barriers, where the barrier width and barrier height of the first and second barrier is $a$, $V_1$, and $b$, $V_2$, respectively. Figure 6(a) and 6(b) shows respectively, the calculated transmission coefficients of electron with an incident energy of $E = 0.1$ eV, at varying ratios of barrier widths ($b/a$) and barrier heights ($V_2/V_1$) between the two constituent barriers. In both cases, very similar dependence of tunneling with the barrier asymmetry is observed, and full transmission ($T(E) = 1$) takes places only for the homo-structured case, where $a = b = 1$ Å, $V_1 = V_2 = 1$ eV. Generally, the total probability of coherent transmission across a rectangular double-barrier may be approximately given by $T \approx 4T_1T_2/(T_1+T_2)^2$ [68], where $T_1$, $T_2$ is the transmission coefficient through each single barrier, respectively. We go further to study the effects of asymmetry on resonance levels, i.e., the quasi-bound states in-between the two barriers (Fig. 1(b)). Figure 6(c) and 6(d) compares respectively, the tunneling spectra for electron and proton across a homo-structured rectangular double-barrier and the corresponding slightly distorted one. It is clearly seen that the distortions in both barrier width and height not only lead to significant changes in the tunneling probability, but also induce small shifts on the resonance peak/level positions.

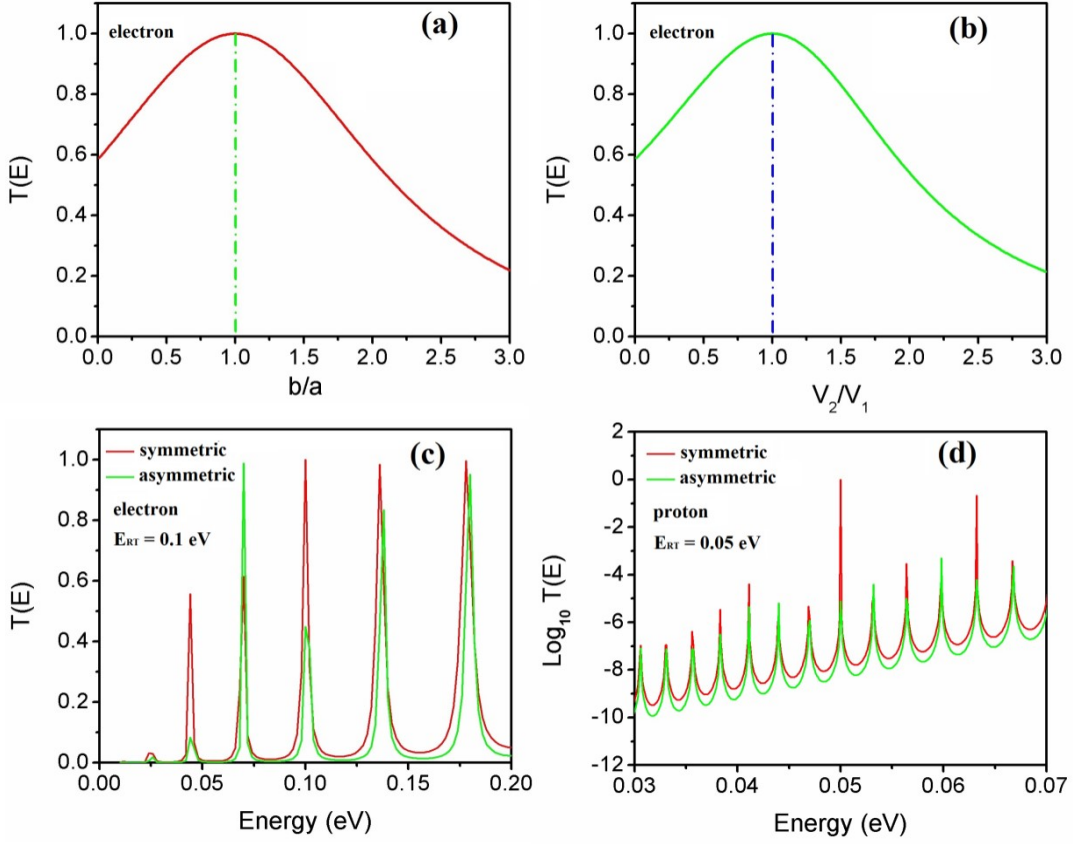

**FIG. 6.** Tunneling properties of electron (**a-c**) and proton (**d**) across homo-structured (symmetric) and hetero-structured (asymmetric) rectangular double-barriers. (**a**) Variation of transmission coefficient $T(E)$ with the barrier width ratio $b/a$, at barrier height $V_1 = V_2 = 1$ eV, and the inter-barrier spacing $w = 109.995$ Å. (**b**) Variation of $T(E)$ with the barrier height ratio $V_2/V_1$, at barrier width $a = b = 1$ Å, and the inter-barrier spacing $w = 109.995$ Å. (**c**) Tunneling spectrum of electron across symmetric and asymmetric double-barriers with given inter-barrier spacing $w = 111.500$ Å, where the parameters $a = b = 2$ Å, $V_1 = V_2 = 1$ eV for the symmetric double-barrier, and the changes of $\delta a = 0.2$ Å, $\delta b = 0.1$ Å, $\delta V_1 = 0.1$ eV, $\delta V_2 = 0.2$ eV for the asymmetric one. (**d**) Similar to (**c**) but for proton tunneling, where the inter-barrier spacing $w = 20.1596$ Å, $a = b = 1$ Å, $V_1 = V_2 = 0.1$ eV for the symmetric one and the changes $\delta a = 0.05$ Å, $\delta V_1 = 0.005$ eV for the asymmetric double-barrier. For panels (**a**)-(**c**) $T(0.1 \text{ eV}) = 1$, and for panel (**d**) $T(0.05 \text{ eV}) = 1$.

For small changes of barrier structures induced by external forces or physical fields such as Casimir forces, finite temperatures, and the gravitational waves, the effects on the tunneling properties may be equivalently ascribed to perturbations on the constituent barriers and be treated within a unified framework. For the special case of resonance at half barrier height ($E_{RT} = 0.5E_b$), like the derivation of Eq. (6) and Eq. (C12) (Appendix C), the dependence of $|(M_{DB})_{11}|^2$ on the more generalized perturbations to double-barrier structure may be similarly deduced by using Taylor series to the second order as below:

$$|(M_{DB})_{11}|^2 \cong 1 + \sinh^2(2ka) \times \left(\frac{kw}{2}\right)^2 \times \left(\frac{\Delta E}{E}\right)^2 + \sinh^4(ka)\left[\left(\frac{\Delta V_1}{V_0}\right)^2 + \left(\frac{\Delta V_2}{V_0}\right)^2\right] +$$
$$\sinh^2(2ka) \times [(k\Delta w)^2 + (k\Delta x_1)^2 + (k\Delta x_2)^2] \qquad (10)$$

where the terms $\Delta E$, $\Delta V_1$, $\Delta V_2$, $\Delta w$, $\Delta x_1$, $\Delta x_2$ are small magnitude of changes in the resonant energy $E$, in the barrier height ($V_0$) of the first and second barrier, in the inter-barrier spacing ($w$), and in the barrier width ($a$) of the first and second barrier, respectively. In the following, we go further to evaluate separately, the effects of Casimir forces and gravitational waves on the tunneling properties across rectangular double-barriers.

Casimir forces are attractive interactions which exist in-between two parallel neutral conducting plates due to the fluctuation of zero-point energy (ZPE) of electromagnetic field in vacuum. The energy change per unit surface area may be calculated as follows [69]: $\Delta E_0(d) = -\frac{\pi^2}{720} \times \frac{\hbar c}{d^3}$, where $d$ is the inter-plate spacing, $\hbar$ is the reduced Planck's constant and $c$ is the speed of light. Casimir forces typically operate in the range of submicron to micron [67, 69]. For a double-barrier with cross-section area $A$, the potential change at a distance of $x$ due to Casimir effect is given by $\Delta V_0(x) = -A \times \frac{\pi^2}{720} \times \frac{\hbar c}{x^3}$, where $w_n \leq x \leq w_n + a + b$. In the our study, the inter-barrier spacing $w_n \gg (a+b)$, and the correction to barrier height can therefore be taken as constant, which is $\Delta V_0(w_n) = -A \times \frac{\pi^2}{720} \times \frac{\hbar c}{w_n^3}$. Given that $A \sim w_n^2$, then $\Delta V_0(w_n) = -\frac{\pi^2}{720} \times \frac{\hbar c}{w_n}$. It follows that $\Delta V_0 \cong 2.7$ meV and 27 meV, respectively, for $w_n = 1\,\mu$m and $0.1\,\mu$m, respectively. The effects of such small magnitude of variations are usually negligible for electron tunneling through the double-barriers studied in this work (Figs. 6(a-c)), except for the situation when the resonance energy ($E_{RT}$) is well below the barrier height (see, e.g., Fig. 3(c)) in which the effects of external perturbations cannot be neglected. Although practically challenging, the low-energy resonant tunneling of electrons points to possible new experimental scheme of detecting the Casimir effect.

Finally, we evaluate the effects of gravitational waves, which are expected to induce expansion or contraction in the geometric size of a double-barrier. Assuming that the strain induced by gravitational waves across the double-barrier region is uniform, then the size changes of a homo-structured rectangular double-barrier as defined in Eq. (10) are given by $\Delta w = h \times w$, $\Delta x_1 = \Delta x_2 = h \times a$, where $h$ is the gravitational-wave strain amplitude projected on the tunneling direction, and $w$ and $a$ are respectively the inter-barrier spacing and barrier width. The peak value of $h$ is adopted in our calculation, which is $1 \times 10^{-21}$ [70]. A passing gravitational wave causes modification on the tunneling probability $T(E)$. In the case of $E_{RT} = 0.5E_b$, the inverse of $T(E)$ may be deduced using Eq. (10) as follows:

$$|(M_{DB})_{11}|^2 = 1 + \sinh^2(2ka) \times h^2 \times [(kw_n)^2 + 2(ka)^2]. \quad (11)$$

The significant change of $T(E)$, in particular $T(E) = 0.5$, i.e., $\delta P = 0.5$ yields $|(M_{DB})_{11}|^2 = 2$, which gives that

$$\sinh^2(2ka) \times h^2 \times [(kw_n)^2 + 2(ka)^2] = 1. \quad (12)$$

The energetic and geometric parameters of a double-barrier for the observation of significant effects induced by gravitational waves are therefore defined by Eq. (12). Let $\chi = 2ka$, recalling that $2kw_n = (2n-1)\pi$ for resonance at half barrier height, Eq. (12) may be rewritten as

$$\sinh(\chi)\sqrt{[(2n-1)\pi]^2 + 2\chi^2} = 2h^{-1}, \quad (13)$$

where $k = \frac{\sqrt{2mE}}{\hbar} = \frac{\sqrt{mE_b}}{\hbar}$. The $n$th root of Eq. (13) $\chi_n = 2ka$, puts constraint on the barrier height ($E_b$) and barrier width ($a$) at inter-barrier spacing $w_n$. For instance, $\chi_1 \cong$ 45.5375 at $w_1 \cong 4.33607172$ Å and $E_b = 1$ eV require that $a \cong 62.85151757$ Å for a significant probability drop ($\delta P = 0.5$) of electron tunneling across a rectangular double-barrier. By contrast, in a general homo-structured rectangular double-barrier, for instance, with $E_b$ = 1 eV, $a$ = 10 Å, and $w_1 \cong$ 4.336 Å at a resonant energy of 0.5 eV, only negligible probability drop ($\delta P = 1.091 \times 10^{-32}$) is expected. Similar situation is found for a given barrier width (e.g., $a$ = 10 Å) where a barrier height ($E_b \cong$ 39.5031326 eV) with ultrahigh accuracy is required to get significant probability drop ($\delta P$ = 0.5). The results are summarized in Table I, for $n$ = 1, 10, and 100. The results indicate that the effects of gravitational waves on resonant tunneling are usually negligible, except for the specially designed double-barriers in which the energetic and geometric parameters with ultrahigh accuracy present.

**Table I.** Energetic and geometric parameters for the resonant tunneling of electrons across homo-structured rectangular double-barriers, and the decrease of transmission probability ($\delta P$) due to the strain ($h = 1 \times 10^{-21}$) induced by gravitational waves. For all the double-barriers, resonance takes place at $E_{RT} = 0.5E_b$. The number of digits are set by $|\Delta w|$, $|\Delta E|$ when $T(E_{RT})$ = 0.5.

| $n$ | $\chi_n$ | $E_b$ (eV) | $w_n$ (Å) | $a$ (Å) | $\delta P$ |
|---|---|---|---|---|---|
| 1 | 45.5375 | 1 | 4.336 | 10 | $1.091 \times 10^{-32}$ |
| | | 1 | 4.3360717180254546576 | 62.8515175683137030659 | 0.5 |
| | | 39.5031326064004559839 | 0.6898913321086561634 | 10 | 0.5 |
| 10 | 45.2686 | 1 | 82.385 | 10 | $1.872 \times 10^{-31}$ |
| | | 1 | 82.3853626424836420483 | 62.4803778905949016575 | 0.5 |
| | | 39.0379762135153995928 | 13.1079353100644677709 | 10 | 0.5 |
| 100 | 43.2978 | 1 | 862.878 | 10 | $1.941 \times 10^{-29}$ |
| | | 1 | 862.8782718870654662168 | 59.7602511637514837162 | 0.5 |
| | | 35.7128761915465986476 | 144.3900008925093061407 | 10 | 0.5 |

The analyses presented above are applicable to the tunneling of a single particle or the situation when the incident particles can be viewed as independent. Nontrivial modifications on the resonant tunneling behavior are expected when the interactions among the incident particles are taken into account. For instance, it is shown experimentally [71, 72] and theoretically [73] that the electrostatic feedback of the space charge dynamically stored in the well of a double-barrier heterostructure produces intrinsic bistability displayed in the *I-V* curve. Generally, the interactions cause splitting of the energy levels of the incident particles from a δ-function like distribution to a broadened energy spectrum, which is physically equivalent to the picture that the tunneling particle induces a (self-consistent) modification of the experienced potential barriers.

Let $\Gamma_n$ being the energy broadening of the *n*th resonance level ($E_n$) in-between the two barriers (Fig. 1(b)), near resonance, the transmission probability may be expressed using the Breit-Wigner formula $T(E) \cong \frac{\Gamma_n^2}{(E-E_n)^2+\Gamma_n^2}$. When $\Gamma_n > \Delta E$, the intrinsic FWHM at energy scale as defined above, the tunneling probability would be enhanced by energy broadening. Equivalently, the inverse of *T(E)*, $|(M_{DB})_{11}|^2$, is reduced. Such an effect due to resonance level broadening may be phenomenologically described as the reduction of *k* value in Eq. (6), by replacing *k* with effective $k' = k - \Delta k$, with $\Delta k = \frac{k}{2}\left|\frac{\Delta E}{E}\right| = \frac{k}{2}\frac{\Gamma_n}{E}$ (Appendix C). For resonance at half barrier height, $2E = E_b$, one has $\Delta k = k\frac{\Gamma_n}{E_b}$, and $k' = k(1-\frac{\Gamma_n}{E_b})$. Then Eq. (6) is rewritten as follows

$$|(M_{DB})_{11}|^2 \cong 1 + \sinh^2\left(2k(1-\frac{\Gamma_n}{E_b})a\right) \times \left(k(1-\frac{\Gamma_n}{E_b})\Delta w\right)^2. \qquad (14)$$

Consequently, the deviation $\Delta w$ which defines the FWHM at length scale is

$$|\Delta w| = \frac{1}{k(1-\frac{\Gamma_n}{E_b})\sinh(2k(1-\frac{\Gamma_n}{E_b})a)}. \qquad (15)$$

For $\Gamma_n = 0$, the particle-particle interactions are negligible and there is no any level broadening, Eq. (15) is simply reduced to Eq. (7), which defines the intrinsic FWHM of resonance at given energy. For any $\Gamma_n > 0$, it is clearly seen that $|\Delta w|$ is enlarged with comparison to the intrinsic value. Depending on the value of $\Gamma_n$, the actual accuracy of position measurement based on the RT of double-barriers could be significantly lower than the ideal case where $\Gamma_n = 0$. For the special case when $\Gamma_n \cong E_b$, $|\Delta w|$ would be infinitely large and *there is no any restriction on the deviation* of inter-barrier spacing. This is understandable since $\Gamma_n \cong E_b$ implies that the resonance levels form a continuous spectrum, and resonance tunneling takes place for any incident energy $E$ ($\leq E_b$). At stationary states, the energy spectrum can be described by the density of states, $D(E)$, which may be numerically calculated using *ab initio* methods like density functional theory (DFT) calculations or quantum Monte Carlo simulations. The tunneling probability $P_t$ is therefore obtained by integration on the spectrum: $P_t = \int T(E) D(E) dE$.

### C. UPPER BOUNDS OF RT BARRIERS SET BY THE PLANCK LENGTH

The critical dependence of the tunneling probabilities on the barrier positions not only demonstrates the crucial role of phase factors, but also points to the possibility of ultrahigh accuracy measurements near resonance. As shown above, the deviation of $|\Delta w| \cong 4.235 \times 10^{-15}$ Å leads to a 50% drop of $P(E)$ of protons across a rectangular double-barrier. Such a deviation is several orders of magnitude below the smallest length scale sensed by LIGO [24, 25, 70]. Even smaller $|\Delta w|$ is expected for heavier particles or larger barriers. As mentioned above, to have measurable RT within some tolerance $\delta P$, an upper bound of deviation from the exact peak position $w_n$ is given by $\Delta w = |w - w_n|$. Suppose that the elementary variation step of distance is $\Delta l$; if the real space is a continuum ($L_{min} = 0$), then $\Delta l$ can be arbitrarily small and in principle $w_n$ can always be reached by a finite number of operations (i.e., $N = [\frac{\Delta w}{\Delta l}]$, for a variation step of $\Delta l$). In this case, the theorem stated above always holds. On the contrary, if a nonzero $L_{min}$ exits ($L_{min} > 0$), to have significant RT, the variation step should satisfy $\Delta w \geq \Delta l \geq L_{min}$. Let $n = \lfloor \text{Log}_{10}(\Delta w) \rfloor$, then it is a feasible choice to set $\Delta l = 10^n$, such that $\Delta w = N \times \Delta l$, with $N$ being an integer and the modulo $\Delta w \mod \Delta l = 0$. Consequently, the existence of a minimum length leads to the inequality of $|\Delta w| \geq L_{min}$, and therefore puts some upper bounds for the particle mass, barrier height and barrier width, above which RT will cease. In the special case of tunneling across rectangular double barriers at $E = 0.5E_b$, realization of RT requires that $|\Delta w| = \frac{1}{k\sinh(2ka)} \geq L_{min}$, which may be rewritten as

$$\chi\sinh(\chi) \leq \frac{2a}{L_{min}}, \qquad (16)$$

where $\chi = 2ka$, $k = \sqrt{\frac{2mE}{\hbar^2}} = \sqrt{\frac{mV_0}{\hbar^2}}$. The upper bound of the term $mV_0$ is therefore determined for a given barrier with $a$. Assuming that the minimum length is identical to the Planck length ($L_{min} = l_P$), the upper bounds of barrier height ($V_{max}$) for electrons and protons are calculated and shown in Fig. 7. At a barrier width of $a$ = 1 Å, 5 Å, and 10 Å, the $V_{max}$ is ~ 5652.72 eV, 239.42 eV, and 61.32 eV for electrons, and is ~ 3.08 eV, 0.13 eV, and 0.03 eV for protons, respectively. It is seen that $V_{max}$ of RT decreases fast with increasing barrier width. For instance, when the barrier width increases to $a$ = 20 and 30 Å, the value of $V_{max}$ is respectively ~ 15.70 and 7.08 eV for electrons, which may be feasible for experimental tests using metal-insulator-metal double barriers. Meanwhile, the same variation trend is found for electrons and protons, with the magnitude of $V_{max}$ being scaled by a factor of $\eta = \frac{m_e}{m_p} \cong \frac{1}{1836}$, where $m_e$ and $m_p$ is respectively the mass of electron and proton. This is due to the conjugate relation that the particle mass times barrier height ($mV_0$) is a constant at fixed barrier width. For the general case of tunneling through arbitrary double barriers, the constraint imposed on the particle mass and barrier size due to a nonzero minimum length is given by (Appendix D):

$$\frac{\hbar}{2\sqrt{2mE}}\sqrt{\frac{1}{R(1+R)}} \times \frac{\delta P}{1-\delta P} \geq L_{min}, \quad (17)$$

where $R = |m_{12}|^2$, and $\delta P$ ($0 < \delta P < 1$) has the same meaning as above. Provided that $L_{min} = l_P$ and $\delta P = 0.5$, the inequality reduces to

$$\frac{\hbar}{2\sqrt{2mE}}\sqrt{\frac{1}{R(1+R)}} \geq l_P. \quad (18)$$

Since the parameter $R$ is generally an increasing function of barrier size (Appendix D), the upper bounds on the barrier size of RT are therefore determined by the Planck length. When the energy broadening due to particle-particle interactions is taken into account, the quantity $k$ is reduced by a factor of $(1 - \frac{\Gamma}{V_0})$, where $\Gamma$ is the level broadening. Based on the analysis above, the product $mV_0$ will be enlarged by a factor of $(1 - \frac{\Gamma}{V_0})^{-2}$.

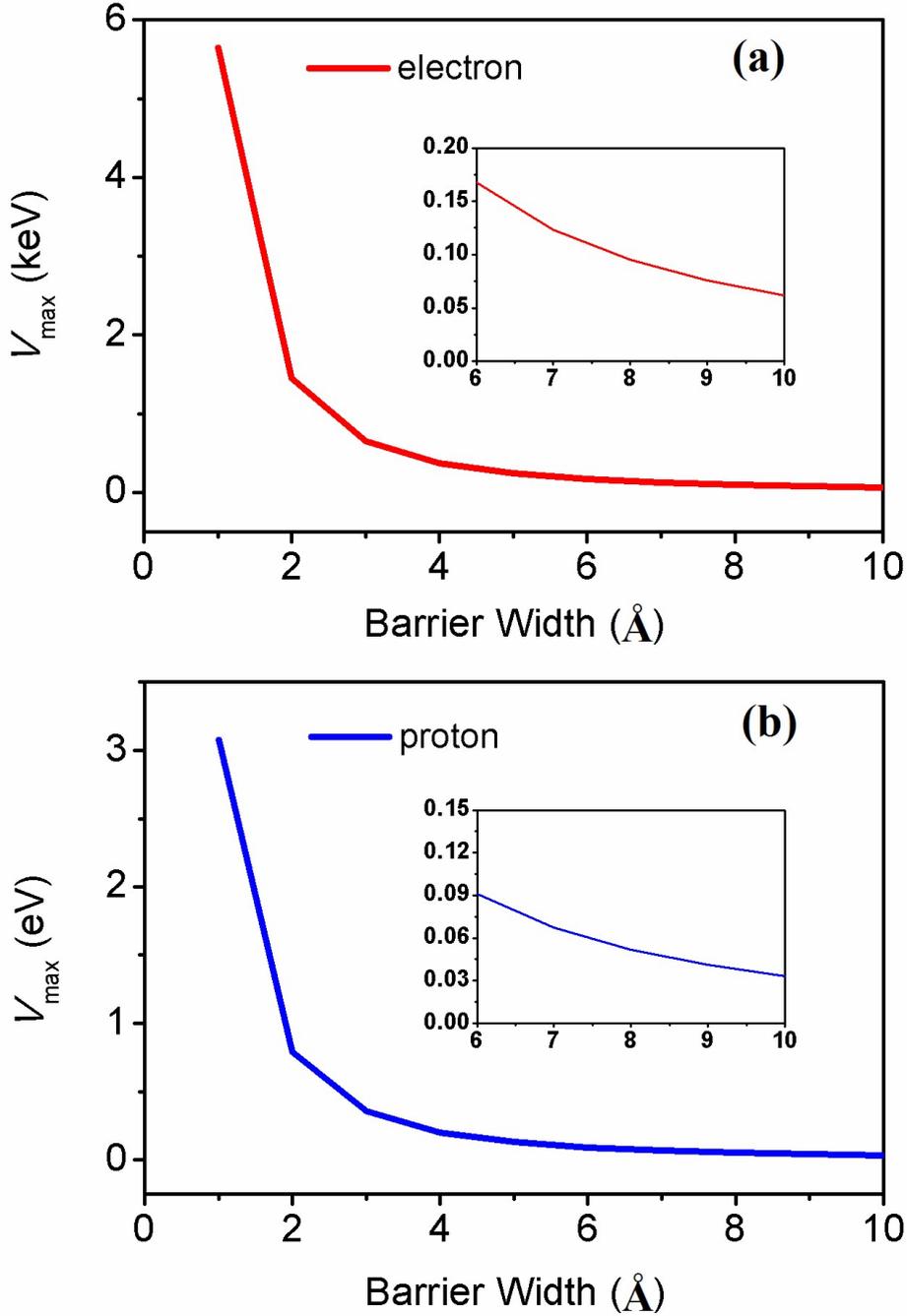

**FIG. 7.** Calculated upper bounds ($V_{max}$) of the barrier height of rectangular double barriers set by the Planck length for electrons **(a)** and protons **(b)**, as a function of barrier width. The values of $V_{max}$ at barrier width of 6-10 Å are highlighted in the insets.

### D. FUNDAMENTAL LIMITS PUT BY THE UNCERTAINTY PRINCIPLE AND POSSIBLE SOLUTION

For a group of incident particles, given that the standard derivation of energy distribution, $\sigma_E$, is approximately the term $\Delta E$ for $\Delta|M_{11}|^2_{\Delta E} = 1$ and $P(E) = 0.5$. The narrow window of energy dispersion implies that the momenta of the particles distribute dominantly within a narrow interval with a small standard derivation ($\Delta p$). As a consequence of the uncertainty principle, the standard derivation of position, $\Delta x$, is expected to be large. Table II lists the energy and momentum broadening, and estimated standard position derivations of protons when $P(E) = 0.5$ at $E = 0.5E_b$, for a number of rectangular double barriers with $E_b$ = 1, 0.5, 0.2, and 0.1 eV, and $w$ ~ 20 Å. It is clearly seen that the energy broadening increases significantly with decreasing barrier height, resulting in reduced standard derivations of position. For $E_b$ = 1 eV, the requirement of ultrahigh monochromaticity of incident energies leads to a very small $\Delta p$ and consequently a quite large $\Delta x$ (~ 1.53×10$^4$ m), which is practically very challenging, if not impossible for experimental tests. Much smaller $\Delta x$ (9.25×10$^{-6}$ m) is found when $E_b$ decreases to 0.1 eV.

It should be stressed here that, the constraint on the standard deviations of particle momentum and position imposed by the uncertainty principle *does not* exclude the possibility that a subgroup of particles with ultrahigh monochromaticity coexist with another subgroup of particles with large energy broadening. The reason is that, by definition, the standard derivation of some physical quantity of a single particle is the statistical average over a large number of events, which may be equivalently evaluated by the statistical results of a large number of identical particles within a small interval of time. Indeed, this is in line with the impossibility of measuring the quantum state of a single system [74]. Therefore, a possible recipe to the practical difficulty of position delocalization is to have much larger standard derivation of kinetic energy distribution than that required for the half drop of $P(E)$, i.e., $\sigma_E \gg \Delta E$, such that the standard derivation of momentum, $\sigma_p$, is much larger than the $\Delta p$ corresponding to $\Delta E$. Consider two microcanonical ensembles containing $N$ weakly interacting identical bosons which follow the kinetic energy distributions $g_1(E)$ and $g_2(E)$ respectively: $N = \int_0^\infty g_1(E)dE = \int_0^\infty g_2(E)dE$. In addition, they have the same averaged kinetic energies: $\langle E \rangle = \int_0^\infty E g_1(E)dE = \int_0^\infty E g_2(E)dE$. The key difference is the standard deviation of kinetic energies: $\sigma_{E2} \gg \sigma_{E1} \sim \Delta E$, i.e., the energy broadening of the first group of particles (distribution described by $g_1(E)$) is approximately the energy deviation for the half drop of $P(E)$, while is much smaller than that of the second group. The distribution function of the mixed $2N$-particle ensemble is $g(E) = g_1(E) + g_2(E)$, with the standard deviation $\sigma_E = \sqrt{\frac{\sigma_{E1}^2 + \sigma_{E2}^2}{2}} \approx \frac{\sigma_{E2}}{\sqrt{2}} \gg \Delta E$. Therefore, mixing of the two groups of identical particles has drastically increased the energy broadening and reduced the position uncertainty. Meanwhile, sufficient number of particles for resonant transmission is maintained. The modifications introduced by the procedure are illustrated in Fig. 8. For the general case of $P(E) = 1 - \delta P$, the energy broadening $\Delta E$ is given by Eq. (9) and can be similarly analyzed. In weakly interacting dilute atomic gases, two-body collisions dominate the interactions which simply exchange particle momenta and therefore keep the kinetic energy distributions unchanged.

The critical dependence of the tunneling behavior with small energy deviations requires that the energies of the incident particles to distribute within a very narrow range, or ideally, with a δ-function-like kinetic energy distribution. In practice, the first group of particles may be prepared using the Bose-Einstein condensates [75-77], in which a large fraction of atoms from a Bose gas occupy the same quantum state, and the momenta of all involved bosons are expected to have *approximately the same value*: Condensation in the momentum space. The second group of particles may be prepared at temperatures slightly above the critical temperature $T_c$ of phase transition from normal states to the new quantum states like superconductivity, superfluidity, or Bose-Einstein condensation. The common feature is that the particles in the first group are of high monochromaticity in the momentum space as well as of high coherence in their wave functions: Both are key factors for the realization of RT in double-barrier systems. At temperatures above $T_c$, the quantum motions of the incident gas atoms may be described in terms of wave packets [78]. As an example, the RT of some typical bosons (Cooper pair of superconducting Nb, $^4$He, $^7$Li, $^{23}$Na, $^{87}$Rb) across rectangular double barriers is studied and the related parameters are presented in Table III. The effects of energy broadening through mixing identical bosons of different ensembles are evidenced by the significantly reduced standard position deviations. As mentioned above, phase coherence of the incident particles plays a key role in the RT process. The quantity characterizing the strength of phase coherence is the coherence length $\xi$, which sets the upper limit for the inter-barrier spacing: $w \leq \xi$. Within the BCS theory, the coherence length of Cooper pairs is given by $\xi = \hbar v_F/\pi\Delta_0$, where $v_F$ is the Fermi velocity, $\Delta_0$ is the order parameter. For a BEC condensate, the coherence length is [79] $\xi \sim 1/\sqrt{na_s}$, where $n$ is the particle density and $a_s$ is the scattering length. In the BCS-BEC crossover region ($\Delta_0 \sim E_F$, the Fermi energy), $\xi \sim 1/k_F$, inverse of the Fermi wave vector, which scales as $1/\sqrt[3]{n}$. In this case, the coherence length of both Cooper pairs and BEC condensates decreases monotonically with particle density. Typical coherence length of Cooper pairs can span from hundreds of angstroms to microns. Nevertheless, preparation of the first group of particles with ultrahigh monochromaticity remains challenging even with state-of-the-art technique. Another challenge to experimental tests may be the acceleration of the condensates as a whole to desired incident velocities while maintaining the states of condensation [80, 81]. Alternatively, the experimentally observed intrinsic bistability in the electron-based asymmetric double-barrier systems [71, 72] may have enlightenment to design similar experimental architectures for testing the existence of $L_{min}$ based on the RT of massive quantum particles (e.g., protons, atoms, molecules). Indeed, atom-based interferometers have demonstrated their feasibility in ultrahigh precision measurements [82-89].

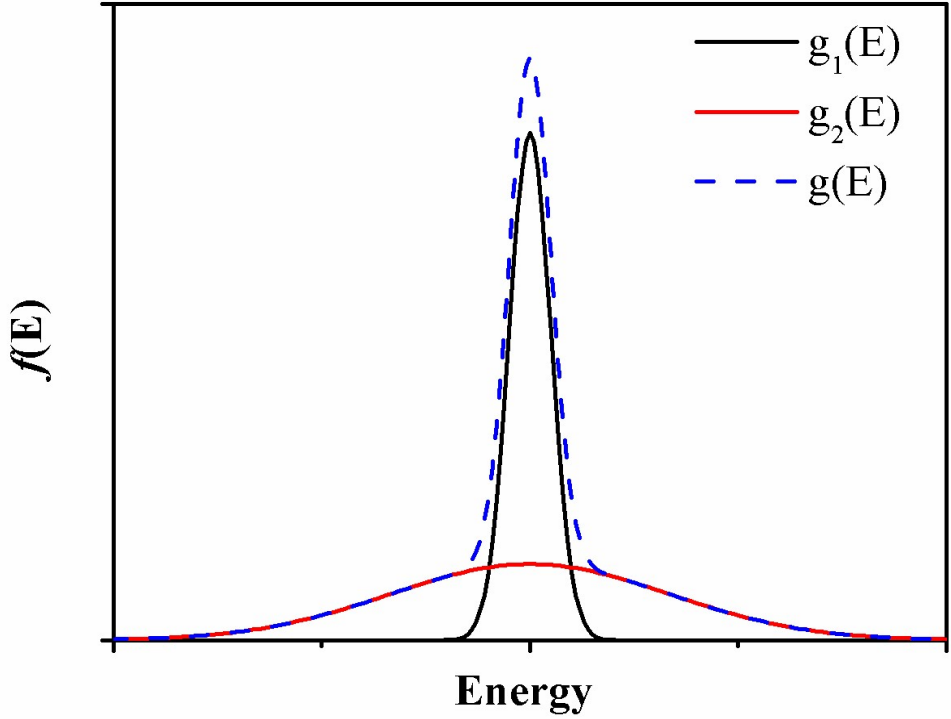

**FIG. 8.** Schematic diagram for the kinetic energy distribution ($f(E)$) of identical particles in microcanonical ensembles: Particle groups of high monochromaticity ($f(E) = g_1(E)$), low monochromaticity ($f(E) = g_2(E)$), and their superposition ($f(E) = g(E) = g_1(E) + g_2(E)$).

Table II. Parameters describing the RT of protons across rectangular double barriers at $E = 0.5E_b$. With the deviation of $\Delta E$ or $|\Delta w|$ from the parameters for resonance, the tunneling probability drops from 1 to 0.5. The corresponding momentum broadening $\Delta p$, and the minimum standard deviation of particle positions $\Delta x_m$ are calculated using the relation $\Delta p \Delta x_m \geq \hbar/2$. In all cases the barrier width $a = 1$ Å.

| $E_b$ (eV) | $w$ (Å) | $\Delta E$ (eV) | $|\Delta w|$ (Å) | $\Delta p$ (kg.m/s) | $\Delta x_m$ (m) |
|---|---|---|---|---|---|
| 1 | 20.137016632763302 | 2.103×10⁻¹⁶ | 4.235×10⁻¹⁵ | 3.443×10⁻³⁹ | 15314.7 |
| 0.5 | 20.17790917547 | 1.320×10⁻¹² | 5.328×10⁻¹¹ | 3.056×10⁻³⁵ | 1.725 |
| 0.2 | 20.1380336 | 2.671×10⁻⁹ | 2.690×10⁻⁷ | 9.778×10⁻³² | 5.392×10⁻⁴ |
| 0.1 | 20.15963 | 1.101×10⁻⁷ | 2.220×10⁻⁵ | 5.700×10⁻³⁰ | 9.251×10⁻⁶ |

**Table III.** Similar to Table II but for the RT of some typical bosons with incident energy $E = 0.5E_b$. In all cases the barrier width $a = 1$ Å, and barrier height $E_b = 0.01V_{max}$, with $V_{max}$ being the upper bound set by the Planck length. The Cooper pairs of electrons are represented by $e^-...e^-$. The energy broadening and resulted uncertainties of momenta and positions of mixed particle groups are displayed in the lower lines of the same columns. The broadening parameter of energy is chosen such that $\sigma_E \gtrsim k_B T_c$, with $k_B$ the Boltzmann constant and $T_c$ the phase transition temperatures.

| Boson | $E_b$ (eV) | $w$ (Å) | $|\Delta w|$ (Å) | $\Delta E$ (eV) | $\Delta p$ (kg.m/s) | $\Delta x_m$ (m) |
|---|---|---|---|---|---|---|
|  |  |  |  | $\sigma_E$ (eV) | $\sigma_p$ (kg.m/s) | $\sigma_x$ (m) |
| $e^-...e^-$ (in Nb) | 28.26 | 6.3439 | 3.16×10⁻³ | 1.41×10⁻¹² | 1.43×10⁻³⁷ | 368.15 |
|  |  |  |  | 1×10⁻³ | 1.02×10⁻²⁸ | 5.19×10⁻⁷ |
| ⁴He | 7.69×10⁻³ | 6.3439 | 3.16×10⁻³ | 3.83×10⁻¹⁶ | 1.43×10⁻³⁷ | 368.15 |
|  |  |  |  | 5×10⁻⁴ | 1.87×10⁻²⁵ | 2.82×10⁻¹⁰ |
| ⁷Li | 4.39×10⁻³ | 6.3439 | 3.16×10⁻³ | 2.19×10⁻¹⁶ | 1.43×10⁻³⁷ | 368.15 |
|  |  |  |  | 1×10⁻¹⁰ | 6.54×10⁻³² | 8.07×10⁻⁴ |
| ²³Na | 1.34×10⁻³ | 6.3439 | 3.16×10⁻³ | 6.67×10⁻¹⁷ | 1.43×10⁻³⁷ | 368.15 |
|  |  |  |  | 1×10⁻¹⁰ | 2.15×10⁻³¹ | 2.45×10⁻⁴ |
| ⁸⁷Rb | 3.53×10⁻⁴ | 6.3439 | 3.16×10⁻³ | 1.76×10⁻¹⁷ | 1.43×10⁻³⁷ | 368.15 |
|  |  |  |  | 1×10⁻¹⁰ | 8.13×10⁻³¹ | 6.49×10⁻⁵ |

### III. CONCLUSIONS

To summarize, we have studied quantum tunneling across double barriers and arrived at a theorem which leads to several physical consequences. First of all, by tuning the inter-barrier spacing, it is possible that low-energy particles penetrate arbitrary finite-sized potential barriers completely via resonant tunneling (RT). This result points to the possibility of significant tunneling of massive quantum particles across large barriers at mild conditions. Secondly, it is possible to construct any desired quasi-bound energy levels within the quantum well formed by the two barriers via adjustment of the inter-barrier spacing. Thirdly, for the RT of quantum particles, it is possible to detect the tiny variations of energy levels and positions of the involved potential barriers with unprecedented accuracy. Finally, the critical dependence on inter-barrier spacing (consequently the phase difference) demonstrates again the vital role of phase factor of wave function, which has manifested itself in some remarkable phenomenon such as the Aharonov-Bohm effect [90].

Demonstration of the above mentioned results involves two key factors: (i) Continuity of the real space and (ii) Energy monochromaticity of the incident particles. The first is determined by whether or not a nonzero minimum length ($L_{min}$) exists, and the second is affected by the uncertainty principle. Provided that $L_{min} = 0$, the distances in real space change continuously and RT can always be realized at given incident energies. On the contrary, the existence of a nonzero $L_{min}$ will set constraints (upper bounds) for the particle mass, barrier height, and barrier width, beyond which no RT is expected. In realistic applications, the energy broadening due to the interactions between the incident particles also puts limit on the accessible accuracy of position determination, and the phase coherence length of the incident particles sets upper bound for the size of double-barriers. Meanwhile, to surmount the practical difficulty (position delocalization of incident particles) owing to the uncertainty principle, we suggest a plausible scheme in which the high- and low-monochromatic beams of identical particle groups are mixed. Potential applications of Bose-Einstein condensates in the scheme are discussed. This work reveals the deep connection between two seemingly different branches of quantum physics: quantum tunneling and quantum gravity, and opens a possible avenue for testing the existence of a minimum length.

## ACKNOWLEDGEMENTS

This work is financially supported by the National Natural Science Foundation of China (No. 11474285, 12074382). The author is grateful to the staff of Hefei Advanced Computing Center for support of supercomputing facilities.

# APPENDIX A: MATRIX ELEMENT $m_{11}$ FOR TUNNELING ACROSS SINGLE RECTANGULAR BARRIER

Within the transfer matrix method, we derive the diagonal matrix element $m_{11}$ that describes the transmission across a single rectangular barrier. For a quantum particle with incident energy $E$ tunneling through a rectangular barrier with the height of $V_0$ and width $a$, the transfer matrix may be given by [56]:

$$M_1 = \frac{1}{2ik}\begin{pmatrix}(ik+\beta)e^{-(ik-\beta)a} & (ik-\beta)e^{-(ik+\beta)a} \\ (ik-\beta)e^{(ik+\beta)a} & (ik+\beta)e^{(ik-\beta)a}\end{pmatrix}\frac{1}{2\beta}\begin{pmatrix}\beta+ik & \beta-ik \\ \beta-ik & \beta+ik\end{pmatrix}$$

$$=\frac{\gamma}{i}\begin{pmatrix}(ik+\beta)^2 e^{-(ik-\beta)a}-(ik-\beta)^2 e^{-(ik+\beta)a} & (\beta^2+k^2)\left(e^{-(ik-\beta)a}-e^{-(ik+\beta)a}\right) \\ (\beta^2+k^2)\left(e^{(ik-\beta)a}-e^{(ik+\beta)a}\right) & (ik+\beta)^2 e^{(ik-\beta)a}-(ik-\beta)^2 e^{(ik+\beta)a}\end{pmatrix}$$

where $k=\sqrt{2mE/\hbar^2}$, $\beta=\sqrt{2m(V_0-E)/\hbar^2}$, $\gamma=\frac{1}{4\beta k}$.

The first diagonal term is: $m_{11}=\frac{\gamma}{i}[(ik+\beta)^2 e^{-(ik-\beta)a}-(ik-\beta)^2 e^{-(ik+\beta)a}]=-i\gamma[(ik+\beta)^2 e^{-(ik-\beta)a}-(ik-\beta)^2 e^{-(ik+\beta)a}]$, which can be reduced to

$m_{11}=-2i\gamma e^{-ika}[(\beta^2-k^2)\sinh(\beta a)+2i\beta k\cosh(\beta a)]$, and finally one has

$$m_{11}=2\gamma e^{-ika}[i(k^2-\beta^2)\sinh(\beta a)+2\beta k\cosh(\beta a)] \qquad (A1)$$

# APPENDIX B: DEDUCTION OF ALTERNATIVE RT CONDITION

In this appendix, we deduce the resonant tunneling (RT) condition for homo-structured rectangular double-barriers.

For a double-barrier (DB) consisting of single rectangular barriers with the height $V_0$ and barrier widths $a$ and $b$, the diagonal element $M_{11}$ of the transfer matrix $M$ may be expressed as follows [57]:

$$|M_{11}|^2 = 1 + \frac{(\beta^2+k^2)^2}{4\beta^2 k^2}[\sinh^2(\beta b)+\sinh^2(\beta a)] + 2\left[\frac{(\beta^2+k^2)^2}{4\beta^2 k^2}\right]^2 \sinh^2(\beta b)\sinh^2(\beta a)$$

$$-\frac{1}{16k^4\beta^4}(\beta^2+k^2)^2 \sinh(\beta b)\sinh(\beta a)\left\{\left[[(\beta^2+k^2)^2-8k^2\beta^2]\cosh\beta(a+b)\right.\right.$$

$$\left.-(\beta^2+k^2)^2\cosh\beta(a-b)\right]\cdot\cos 2[k(b+w)-ka]$$

$$-4k\beta(\beta^2-k^2)\sinh\beta(a+b)\cdot\sin 2[k(b+w)-ka]\bigg\}. \qquad (B1)$$

where $k = \sqrt{2mE/\hbar^2}$, $\beta = \sqrt{2m(V_0 - E)/\hbar^2}$, $E$ is energy of the incident particle.

In the case of homo-structured DB, $a = b$, then

$$|M_{11}|^2 = 1 + \frac{(\beta^2+k^2)^2}{4\beta^2 k^2} \times [2\sinh^2(\beta a)] + 2\left[\frac{(\beta^2+k^2)^2}{4\beta^2 k^2}\right]^2 \sinh^4(\beta a) - \frac{1}{16k^4\beta^4}(\beta^2 +$$

$$k^2)^2 \sinh^2(\beta a) \times \{[[(\beta^2+k^2)^2 - 8k^2\beta^2]\cosh(2\beta a) - (\beta^2+k^2)^2]\cos(2kw) -$$

$$4k\beta(\beta^2 - k^2)\sinh(2\beta a) \cdot \sin(2kw)\}. \tag{B2}$$

For a given $E$, $|M_{11}|^2 = \frac{1}{T(E;w)}$ is the function of inter-barrier spacing $w$, the minimum of $|M_{11}|^2$ gives the maximum of transmission coefficient $T(E; w)$, i.e., resonant tunneling (RT). The condition of RT can be established by $\frac{\partial}{\partial w}|M_{11}|^2 = 0$. It follows that,

$$[[(\beta^2+k^2)^2 - 8k^2\beta^2]\cosh(2\beta a) - (\beta^2+k^2)^2] \times (-2k)\sin(2kw) -$$

$4k\beta(\beta^2 - k^2)\sinh(2\beta a) \times (2k)\cos(2kw) = 0$, and consequently

$$\tan(2kw) = \frac{4k\beta(\beta^2-k^2)\sinh(2\beta a)}{(\beta^2+k^2)^2 - [(\beta^2+k^2)^2 - 8k^2\beta^2]\cosh(2\beta a)} \tag{B3}$$

By dividing the term $\beta^2 k^2$ in both numerator and denominator, Eq. (B3) changes to

$$\tan(2kw) = \frac{4\left(\frac{\beta}{k}-\frac{k}{\beta}\right)\sinh(2\beta a)}{\left(\frac{\beta}{k}+\frac{k}{\beta}\right)^2 - \left[\left(\frac{\beta}{k}+\frac{k}{\beta}\right)^2 - 8\right]\cosh(2\beta a)} \equiv \frac{4\delta \sinh(2\beta a)}{(\delta^2+4)-(\delta^2-4)\cosh(2\beta a)} =$$

$$\frac{\delta \sinh(2\beta a)}{\left(1+\frac{1}{4}\delta^2\right)+\left(1-\frac{1}{4}\delta^2\right)\cosh(2\beta a)}, \text{ where } \delta \equiv \left(\frac{\beta}{k} - \frac{k}{\beta}\right).$$

Recalling that $\sinh(2\beta a) = 2\sinh(\beta a)\cosh(\beta a)$, $\cosh(2\beta a) = 2\cosh^2(\beta a) - 1$, one has

$$\tan(2kw) = \frac{2\delta \sinh(\beta a)\cosh(\beta a)}{\left(1+\frac{1}{4}\delta^2\right)+(1-\frac{1}{4}\delta^2)(2\cosh^2(\beta a)-1)} = \frac{2\delta \sinh(\beta a)\cosh(\beta a)}{\frac{1}{2}\delta^2+(1-\frac{1}{4}\delta^2)\times\cosh^2(\beta a)}, \text{ which can be}$$

reduced to

$$\tan(2kw) = \frac{\delta \tanh(\beta a)}{\left(1-\frac{1}{4}\delta^2\right)+\frac{1}{4}\frac{\delta^2}{\cosh^2(\beta a)}} = \frac{\delta \tanh(\beta a)}{\left(1-\frac{1}{4}\delta^2\right)+\frac{\delta^2}{4}\text{sech}^2(\beta a)}. \tag{B4}$$

Using the equality $\text{sech}^2(\beta a) = 1 - \tanh^2(\beta a)$, Eq.(B4) is finally reduced to

$$\tan(2kw) = \frac{\delta \tanh(\beta a)}{1 - \frac{\delta^2}{4}\tanh^2(\beta a)} \tag{B5}$$

# APPENDIX C: DEPENDENCE OF TUNNELING ON SMALL POSITION AND ENERGY CHANGES

In this appendix, we deduce the mathematical expressions describing the dependence of squared norm of diagonal transfer matrix element, $|(M_{DB})_{11}|^2$, with respect to slight deviations from the peak positions and incident energies at resonant tunneling (RT), for the special case when the incident energy is half the barrier height ($V_0$) of a homo-structured rectangular double-barrier (width of single barrier: $a$). The inverse of $|(M_{DB})_{11}|^2$ then describes the dependence of tunneling behavior on small position and energy changes.

In general, $|(M_{DB})_{11}|^2 \equiv f(E;w)$, is the function of incident energy $E$ and inter-barrier spacing $w$. At the vicinity of RT, the function $f(E;w)$ can be expressed as functions of small deviations from RT parameters using the Taylor series, by considering the fact that $|(M_{DB})_{11}|^2 = 1$ and $\left(\frac{\partial f}{\partial w}\right) = 0$, $\left(\frac{\partial f}{\partial E}\right) = 0$ at the RT point.

**I.** For constant $E$, the dependence on deviation ($\Delta w$) from the RT positions ($w_n$) is

$$|(M_{DB})_{11}|^2 \equiv f(E;w) \cong 1 + \frac{1}{2}\left(\frac{\partial^2 f}{\partial w^2}\right) \times (\Delta w)^2 \equiv 1 + \Delta|M_{11}|^2_{\Delta w} \tag{C1}$$

Using the expressions for rectangular double barriers (Appendix B), one has

$$\frac{\partial f}{\partial w} = -\left(\frac{1}{16\beta^4 k^4}\right)(\beta^2 + k^2)^2 \sinh^2(\beta a)[g(\beta,k) \times (-2k) \times \sin(2kw) + h(\beta,k) \times (2k) \times \cos(2kw)], \tag{C2}$$

$$\frac{\partial^2 f}{\partial w^2} = \frac{(\beta^2 + k^2)^2 \sinh^2(\beta a)}{4\beta^4 k^2}[g(\beta,k)\cos(2kw) + h(\beta,k)\sin(2kw)], \tag{C3}$$

where $k = \sqrt{2mE/\hbar^2}$, $\beta = \sqrt{2m(V_0 - E)/\hbar^2}$,

$g(\beta,k) \equiv [(\beta^2 + k^2)^2 - 8\beta^2 k^2]\cosh(2\beta a) - (\beta^2 + k^2)^2$, $h(\beta,k) = -4\beta k(\beta^2 - k^2)\sinh(2\beta a)$.

The condition $\left(\frac{\partial f}{\partial w}\right) = 0$ gives that $g(\beta,k)\sin(2kw) = h(\beta,k)\cos(2kw)$, and then

$$\tan(2kw) = \frac{h(\beta,k)}{g(\beta,k)} \tag{C4}$$

Using Eq. (C4), $\frac{\partial^2 f}{\partial w^2} = \frac{(\beta^2+k^2)^2 \sinh^2(\beta a)}{4\beta^4 k^2} h(\beta,k)\sin(2kw)[\cot^2(2kw) + 1]$, and then

$$\frac{\partial^2 f}{\partial w^2} = \frac{(\beta^2+k^2)^2 \sinh^2(\beta a)}{4\beta^4 k^2} \times \frac{h(\beta,k)}{\sin(2kw)} \tag{C5}$$

For rectangular double barriers, we have the general relation $2kw = (2n-1)\pi - 2\alpha$, and $\alpha = \arctan[\frac{(k^2-\beta^2)}{2\beta k}\tanh(\beta a)]$.

Consequently,

$$\sin(2kw) = \sin(2\alpha) = \frac{2\tan\alpha}{1+\tan^2\alpha} = \frac{2\frac{(k^2-\beta^2)}{2\beta k}\tanh(\beta a)}{1+\tan^2\alpha}.$$

Finally,

$$\frac{\partial^2 f}{\partial w^2} = \frac{(\beta^2+k^2)^2 \sinh^2(\beta a)}{4\beta^4 k^2} \frac{h(\beta,k)}{\sin(2kw)} = \frac{(\beta^2+k^2)^2 \sinh^2(2\beta a)}{2\beta^2}(1+\tan^2\alpha) \tag{C6}$$

It is clear that $\frac{\partial^2 f}{\partial w^2} > 0$ holds for all allowed incident energies *E, which proves that the term* $|(M_{DB})_{11}|^2$ *arrives at its minimum and its reciprocal gives the maximum of transmission probability, i.e.,* 1.

When the incident energy is half the barrier height, $\beta = k$, we have $\alpha = 0$, and $\frac{\partial^2 f}{\partial w^2} = 2k^2 \sinh^2(2ka)$, and therefore

$$|(M_{DB})_{11}|^2 \cong 1 + \frac{1}{2}\left(\frac{\partial^2 f}{\partial w^2}\right) \times (\Delta w)^2 = 1 + \sinh^2(2ka) \times (k\Delta w)^2 \tag{C7}$$

**II.** For constant $w$, the dependence on deviation ($\Delta E$) from the RT energies ($E_{RT}$) is

$$|(M_{DB})_{11}|^2 \equiv f(E;w) \cong 1 + \frac{1}{2}\left(\frac{\partial^2 f}{\partial E^2}\right) \times (\Delta E)^2 \equiv 1 + \Delta|M_{11}|^2_{\Delta E} \tag{C8}$$

Compared to $\left(\frac{\partial^2 f}{\partial w^2}\right)$, computation of $\left(\frac{\partial^2 f}{\partial E^2}\right)$ is much more complicated. Alternatively, we directly consider the dependence of $|(M_{DB})_{11}|^2$ with energy deviation ($\Delta E$) to the second order. For the special situation $\beta = k$, the mathematical expression of $|(M_{DB})_{11}|^2$ is reduced to

$$|(M_{DB})_{11}|^2 = 1 + 2\sinh^2(ka) + 2\sinh^4(ka) + \sinh^2(ka)[\cosh(2ka) + 1] \times \cos(2kw) \tag{C9}$$

Recalling that $2kw = (2n-1)\pi$ for $\beta = k$, the term $\cos(2kw)$ may be expressed by Taylor series around the RT point with respect to $\Delta k$ to the second order:

$$\cos(2kw) \cong -1 + \frac{1}{2}(2w)^2(\Delta k)^2 = -1 + 2(w\Delta k)^2 \qquad (C10)$$

Substitution of $\cos(2kw)$ with Eq. (C10) leads to the following

$$|(M_{DB})_{11}|^2 \cong 1 + \sinh^2(2ka)(w\Delta k)^2 \qquad (C11)$$

Using $k = \sqrt{2mE/\hbar^2}$, and then $\Delta k = \frac{\Delta E}{2\sqrt{E}}\sqrt{2m/\hbar^2} = \frac{k}{2} \times \frac{\Delta E}{E}$, one finally arrives at

$$|(M_{DB})_{11}|^2 \cong 1 + \sinh^2(2ka)(\frac{kw}{2})^2(\frac{\Delta E}{E})^2 \qquad (C12)$$

# APPENDIX D: GENERALIZED CONSTRAINTS ON BARRIER SIZE DUE TO A MINIMUM LENGTH

In this appendix, we deduce the constraint on the barrier size (barrier height, barrier width) for effective resonant tunneling (RT) at the presence of a nonzero minimum length ($L_{min}$). As defined above, effective RT implies that giving the deviation $\Delta w$ when $|w - w_n| \leq \Delta w$, the inequality $T(E; w) \geq 1 - \delta P$ holds, where $\delta P$ ($0 < \delta P < 1$) is the tolerance of decrease in tunneling probability at which significant tunneling is measurable. Based on the proof of the theorem, we have

$$T_{DB}(E; w) = \frac{1}{|(M_{DB})_{11}|^2} = \frac{1}{|e^{i\theta}[1+R(e^{-i(\phi+\theta)}+1)]|^2} = \frac{1}{|1+R(e^{-i(\phi+\theta)}+1)|^2} \qquad (D1)$$

Then $\Delta w$ is determined by the equality as follows

$$\frac{1}{|1+R(e^{-i(\phi+\theta)}+1)|^2} = 1 - \delta P. \qquad (D2)$$

Equivalently,

$$\left|1 + R(e^{-i(\phi+\theta)} + 1)\right|^2 = \frac{1}{1-\delta P} . \qquad (D3)$$

It follows that

$$|1 + R + R(\cos(\phi+\theta) - i\sin(\phi+\theta))|^2 = \frac{1}{1-\delta P} , \qquad (D4)$$

$$(1 + R + R\cos(\phi+\theta))^2 + R^2\sin^2(\phi+\theta) = \frac{1}{1-\delta P} , \qquad (D5)$$

$$(1+R)^2 + R^2 + 2(1+R)R\cos(\phi+\theta) = \frac{1}{1-\delta P} , \qquad (D6)$$

and then reduces to

$$2R(1+R)\left[1+\cos(\phi+\theta)\right] = \frac{\delta P}{1-\delta P}. \tag{D7}$$

Consequently, we arrive at

$$\cos(\phi+\theta) = -1 + \frac{1}{2R(1+R)} \times \frac{\delta P}{1-\delta P}. \tag{D8}$$

For a homo-structured double-barrier system, the two parameters $\theta = \arg(m_{11}{}^2)$, and $R = |m_{12}|^2$ are solely determined by a single barrier $V(x)$. The tunable parameter is $\phi = 2k(a+w)$, via variation of the inter-barrier spacing $w$ by a small magnitude of $\Delta w$. At the vicinity of RT, $|\Delta w| \ll w_n$. Around $\phi + \theta = (2n-1)\pi$, i.e., the RT points, expansion of $\cos(\phi+\theta)$ using Taylor series to the 2nd order, we have

$$\cos(\phi+\theta) \cong -1 + \frac{1}{2}(\Delta\phi)^2, \tag{D9}$$

where $\Delta\phi = \pm 2k\Delta w$. Comparison of Eq. (D8) and (D9) gives that

$$2k\Delta w = \sqrt{\frac{1}{R(1+R)} \times \frac{\delta P}{1-\delta P}}. \tag{D10}$$

Finally, we get

$$\Delta w = \frac{1}{2k}\sqrt{\frac{1}{R(1+R)} \times \frac{\delta P}{1-\delta P}}. \tag{D11}$$

To achieve effective RT, the existence of $L_{min}$ requires that

$$\Delta w = \frac{1}{2k}\sqrt{\frac{1}{R(1+R)} \times \frac{\delta P}{1-\delta P}} \geq L_{min} \tag{D12}$$

For a single barrier $V(x)$, the reflection coefficient is given by [59-62]

$$|r|^2 = \frac{|m_{12}|^2}{|m_{11}|^2} = RT_1(E) = R|t|^2, \tag{D13}$$

where $R = |m_{12}|^2$, and $T_1(E) = |t|^2$ is the transmission coefficient across $V(x)$ at energy $E$. Conservation of probability current gives that $|r|^2 + |t|^2 = 1$. Qualitatively, $|r|^2$ increases with barrier width $a$ and barrier height $E_b$, which indicates that *R is the increasing function of barrier size* parameters $a$ and $E_b$: $R = R(a, E_b)$. Larger barrier size results in larger value of $R$. Substitution of $k$ with $\frac{\sqrt{2mE}}{\hbar}$, the inequality (D12) is therefore

$$\frac{\hbar}{2\sqrt{2mE}}\sqrt{\frac{1}{R(1+R)} \times \frac{\delta P}{1-\delta P}} \geq L_{min} \tag{D14}$$

This is the constraint imposed on the particle mass, barrier height, and barrier width due to the minimum length.

In the case $\delta P = 0.5$, FWHM ($= 2\Delta w$) is obtained. Given that $L_{min} = l_P$, we have

$$\frac{\hbar}{2\sqrt{2mE}}\sqrt{\frac{1}{R(1+R)}} \geq l_P \qquad (D15)$$

For a fixed particle mass *m* and incident energy *E*, the inequality (D15) sets upper bounds on $R$ and consequently the upper bounds for barrier size of *V(x)*: the barrier width *a* and barrier height $E_b$.

Furthermore, we can derive the constraint on the broadening of incident energy by using Eq. (D9). In this case, $w$ ($= w_n$) and *a* are fixed, $\Delta\phi = 2\Delta k(a + w)$. Using $k = \frac{\sqrt{2mE}}{\hbar}$, we have $\Delta k = \frac{k}{2} \times \frac{\Delta E}{E}$, then $2\Delta k = k \times \frac{\Delta E}{E}$, and $\Delta\phi = k(a + w) \times \frac{\Delta E}{E}$. It follows that

$$(\Delta\phi)^2 = \frac{1}{R(1+R)} \times \frac{\delta P}{1-\delta P} \quad . \qquad (D16)$$

Then

$$\Delta\phi = k(a + w) \times \left|\frac{\Delta E}{E}\right| = \sqrt{\frac{1}{R(1+R)} \times \frac{\delta P}{1-\delta P}} \quad . \qquad (D17)$$

Finally we have

$$\left|\frac{\Delta E}{E}\right| = \frac{1}{k(a+w)}\sqrt{\frac{1}{R(1+R)} \times \frac{\delta P}{1-\delta P}} \qquad (D18)$$